\documentclass[%
 aip,
 amsmath,amssymb,
 reprint,%
]{revtex4-1}

\usepackage{graphicx}
\usepackage{dcolumn}
\usepackage{bm}

\usepackage[utf8]{inputenc}
\usepackage[T1]{fontenc}
\usepackage{mathptmx}
\usepackage{etoolbox}

\makeatletter
\def\@email#1#2{%
 \endgroup
 \patchcmd{\titleblock@produce}
  {\frontmatter@RRAPformat}
  {\frontmatter@RRAPformat{\produce@RRAP{*#1\href{mailto:#2}{#2}}}\frontmatter@RRAPformat}
  {}{}
}%
\makeatother
\begin{document}

\preprint{AIP/123-QED}

\title[Deep Structured Neural Networks]{Deep Structured Neural Networks for Turbulence Closure Modelling}
\author{R. McConkey}
 \affiliation{Department of Mechanical and Mechatronics Engineering, University of Waterloo
 }
 \email{rmcconke@uwaterloo.ca}
\author{E. Yee}%
 
\affiliation{Department of Mechanical and Mechatronics Engineering, University of Waterloo
}

\author{F.S. Lien}

\affiliation{%
Department of Mechanical and Mechatronics Engineering, University of Waterloo
}%

\date{\today}

\begin{abstract}
Despite well-known limitations of Reynolds-averaged Navier-Stokes (RANS) simulations, this methodology remains the most widely used tool for predicting many turbulent flows, due to computational efficiency. Machine learning is a promising approach to improve the accuracy of RANS simulations. One major area of improvement is using machine learning models to represent the complex relationship between the mean flow field gradients and the Reynolds stress tensor. In the present work, modifications to improve the stability of previous optimal eddy viscosity approaches for RANS simulations are presented and evaluated. The optimal eddy viscosity is reformulated with a non-negativity constraint, which promotes numerical stability. We demonstrate that the new formulation of the optimal eddy viscosity improves the conditioning of the RANS equations for a periodic hills test case. To demonstrate the suitability of this proportional/orthogonal tensor decomposition for use in a physics-informed data-driven turbulence closure, we use two neural networks (structured on this specific tensor decomposition which is incorporated as an inductive bias into the network design) to predict the newly reformulated linear and non-linear parts of the Reynolds stress tensor. Injecting these network model predictions for the Reynolds stresses into a RANS simulation improves predictions of the velocity field, even when compared to a sophisticated (state of the art) physics-based turbulence closure model. Finally, we apply SHAP (SHapley Additive exPlanations) values to obtain insights from the learned representation for the inner workings of the neural network used to predict the optimal eddy viscosity from the input feature data.
\end{abstract}

\maketitle
\section{\label{sec:introduction}Introduction}


In eddy viscosity modelling, the specific relationship between the Reynolds stress tensor and the mean velocity gradient tensor is the primary subject of interest. Many relationships have been proposed, from simple mixing length models to the sophisticated elliptic blending $k$-$\varepsilon$ model. Despite the well-known limitations of the eddy viscosity models for turbulent flows, most industrial simulations still rely on these types of turbulence closures within the Reynolds-averaged Navier-Stokes (RANS) framework because of their computational simplicity and efficiency \cite{Witherden2017}. While the inherent assumption of a local dependence in eddy viscosity modelling is nearly impossible to avoid, improving the functional relationship between the Reynolds stress tensor and the mean velocity gradient tensor remains critically important for accurate industrial simulations \cite{CFD2030}. 


Traditionally, turbulence closure models are formulated using heuristic and physics-based arguments. Common examples include the analogy drawn between turbulent stresses and viscous stresses, or the model transport equations for turbulent kinetic energy (TKE) $k$ and a length-scale determining variable such as $\varepsilon$ (TKE dissipation rate). However, a new approach in turbulence modelling is the physics-\textit{informed} data-driven approach \cite{Duraisamy2020}, which leverages the capabilities of machine learning. Rather than simplifying the underlying physics to close the RANS equations, the functional relationship between the mean flow gradients and Reynolds stresses is determined from high-fidelity datasets. Machine learning is well-suited to ``discover'' this (complex) functional relationship using a data-driven approach.

Within the framework of the application of machine learning for data-driven closure modelling, a wide range of methods have been presented and nearly all of them have demonstrated promising results for simple flows~\cite{Duraisamy2019a,Brunton2020,Kutz2017,Wang2017}. Several machine learning and injection frameworks have been proposed~\cite{Duraisamy2021,Zhu2020,Chang2018,Jiang2021}, including open-loop~\cite{Bhushan2021} and iterative frameworks~\cite{Liu2021}. These methods have been demonstrated to improve RANS results in canonical flows~\cite{Srinivasan2019,Yin2020,Fang2020,Zhang2019b}, flow over airfoils~\cite{Zhu2019,Matai2019}, combustion~\cite{Nikolaou2020}, and transonic flows~\cite{Tan2021}. Several investigations~\cite{Kaandorp2018,Kaandorp2020,Song2019,Zhang2019a} have used machine learning to determine the coefficients of a tensor basis expansion for the anisotropy tensor, based on the seminal work of Ling {\em et al.}~\cite{Ling2016}. Other frameworks include the optimal eddy viscosity approach proposed by Wu {\em et al.}~\cite{Wu2018}, or the Reynolds force vector approach advocated by Cruz {\em et al.}~\cite{Cruz2019}. However, the issue of ill-conditioning was recently shown by Brener {\em et al.}~\cite{Brener2021} to affect a number of these modelling frameworks. When the RANS equations are ill-conditioned, small errors in the Reynolds stress tensor can be amplified, resulting in large errors in the prediction of the mean velocity field~\cite{Wu2019}. This error amplification is a major issue for physics-informed data-driven closures, as some amount of model error in predicting the Reynolds stress tensor is expected. Brener {\em et al.}~\cite{Brener2021} showed that the optimal eddy viscosity approach proposed by Wu {\em et al.}~\cite{Wu2018} is an effective strategy to address the ill-conditioning problem.


While previous investigations have presented optimal eddy viscosity-based frameworks that improve the conditioning of the RANS equations, these formulations provide no guarantee of a non-negative eddy viscosity. For practical purposes, the eddy viscosity should always be non-negative---a negative eddy viscosity prediction has the potential to destabilize convergence in an iterative RANS solver. A new formulation of the optimal eddy viscosity is proposed in the present work, which guarantees a non-negative optimal eddy viscosity. The conditioning of the new formulation is analyzed, and the suitability for use in a physics-informed data-driven turbulence closure model is assessed by using a neural network to predict this optimal eddy viscosity. An additional neural network is used to predict the remaining (non-linear) portion of the Reynolds stress tensor. The prediction of the velocity field is greatly improved after injecting the neural network-predicted Reynolds stress tensor into the mean momentum equations. 


The present work is the first to propose a stability constraint on the optimal eddy viscosity and to implement this constraint into a neural network. Along with proposing a neural network architecture for predicting the optimal eddy viscosity, the present work also aims to demonstrate the ability of deep learning augmentation to improve the results of a sophisticated physics-based turbulence closure model. While previous studies have improved results from the simpler $k$-$\varepsilon$ \cite{Ling2016} and $k$-$\omega$ models \cite{Kaandorp2020}, the model augmented in the present work is the $k$-$\varepsilon$-$\phi_t$-$f$ model \cite{Hanjalic2004},
which is an improved reformulation of the $v2$-$f$ model \cite{Durbin1991}. This model contains three turbulent transport equations for $k$, $\varepsilon$, and $\phi_t$, along with an elliptic equation for the damping variable $f$.


The present work is structured as follows: in Section~\ref{sec:methodology}, we motivate and discuss the proposed optimal eddy viscosity formulation, and injection process. The choice of the network topology and the associated hyperparameter tuning for the customization of deep neural networks for the task of turbulence closure modelling is then described in Section~\ref{sec:deeplearning}. Section~\ref{sec:results} presents the model predictions for an unseen test case. An \textit{a priori} analysis of the model predictions is given in Section~\ref{sec:apriori}. After injecting the model predictions into a RANS solver, the converged mean fields are analyzed in an \textit{a posteriori} sense in Section~\ref{sec:aposteriori}. Having demonstrated the accuracy improvements gained through machine learning augmentation, Section~\ref{sec:interpret} presents an interpretation and discussion of the data-driven closure. Conclusions and recommendations for further study are given in Section~\ref{sec:conclusion}.

\section{Methodology}\label{sec:methodology}

The Reynolds-averaged continuity and momentum equations for an incompressible, isothermal, steady-state mean velocity field $\vec U = (U,V,W)$ and pressure field $p$ is given by
\begin{eqnarray}
\nabla \cdot \vec{U} = 0\ ,\\
\nabla\cdot (\vec{U}\vec{U}) = - \nabla p + \nu \nabla^2 \vec{U} - \nabla \cdot \tau\ ,\label{eq:momentum}
\end{eqnarray}
where $\nu$ is the molecular kinematic viscosity of the fluid and $\tau \equiv \overline{\vec{u}'\vec{u}'}$ is the Reynolds stress tensor. Here, mean values are denoted using an overbar and $\vec{u}' \equiv (\vec{u} - \vec{U})$ is the fluctuating velocity ($\vec{u}$ is the total instantaneous turbulent velocity).

These equations are hereinafter referred to as the RANS equations. The unclosed term (Reynolds stress tensor), $\tau$, is the subject of turbulence closure modelling. The common modelling approach is to approximate the anisotropic part of $\tau$  analogously to the viscous stress term, so
\begin{equation}
a \equiv  \tau - \frac{1}{3}\text{tr}(\tau)I \approx -2 \nu_t S \ , \label{eq:anisotropy}
\end{equation}
where $a$ is the anisotropy tensor and $I$ is the identity tensor. Equation~(\ref{eq:anisotropy}) is known as the eddy viscosity approximation, where $\nu_t$ is the eddy viscosity, and $S$ is the mean strain-rate tensor (symmetric part of the mean velocity gradient tensor) defined by
$$S \equiv \tfrac{1}{2}\left(\nabla \vec{U} + \nabla \vec{U}^\text{T}\right)\ ,$$
where the superscript $T$ denotes matrix transposition.


This approximation for $\tau$ permits the eddy viscosity to be treated implicitly, and added to the molecular viscosity $\nu$ in Eq.~(\ref{eq:momentum}). Numerically, this has a stabilizing effect and therefore this approach is widely used in RANS modelling. Additionally, the term $\nabla \cdot \tfrac{1}{3}\text{tr}(\tau)I$ (which is equal to $\nabla \cdot \tfrac{2}{3}kI$) is typically absorbed into the isotropic term $\nabla p$. The resulting isotropic term becomes the gradient of the ``modified pressure'' given by $p' = p + \tfrac{2}{3}k$. 

While this approximation for $\tau$ is common in traditional turbulence modelling, the approach in data-driven turbulence modelling has not been uniform. Liu {\em et~al.}~\cite{Liu2021} and Brener {\em et al.}~\cite{Brener2021} used the eddy viscosity approximation as follows:
\begin{eqnarray}
\tau \approx -2 \nu_t S \label{eq:eddyviscosity_tau}. 
\end{eqnarray}
However, a deficiency with the approximation given by Eq.~(\ref{eq:eddyviscosity_tau}) is that $\tau$ (whose non-zero trace is $2k$) can never be completely aligned with $S$ (whose trace is necessarily zero owing to the incompressibility of the flow). Therefore, when formulating an ``optimal eddy viscosity approach'' (Section \ref{sec:optimal}), it is more appropriate to use the more conventional approximation given in Eq.~(\ref{eq:anisotropy}).

\subsection{Optimal eddy viscosity}\label{sec:optimal}
Within the commonly used eddy viscosity approach, closing the RANS equations is achieved using a single constitutive coefficient $\nu_t$. However, calculating $\nu_t$ in a way that minimizes the error in the mean flow field is a complex task, and has led to the development of hundreds of RANS closure models. In the present work, we propose to use a neural network to directly estimate $\nu_t$, effectively eliminating the need for any additional turbulent scalar transport equations. This approach was also applied by Wu {\em et al.}~\cite{Wu2018} and Liu {\em et al.}~\cite{Liu2021}.

The optimal eddy viscosity $\nu^*_t$ is obtained by minimizing the error in the approximation of the anisotropy tensor as being directly proportional to the mean strain-rate tensor. More specifically, the optimal eddy viscosity is obtained as the solution of the following least-squares approximation problem:
\begin{eqnarray}
\nu^*_t = \text{arg min}_{\nu_t} \| a - (-2\nu_t S) \|\ ,\label{eq:leastsquares}
\end{eqnarray}
where $\nu^*_t$ is the optimal eddy viscosity in a least-squares sense ($\|\ \cdot\ \|$ denotes the Euclidean norm). Equation~(\ref{eq:leastsquares}) has the following closed-form analytical solution: 
\begin{eqnarray}
\nu^*_t= -\dfrac{1}{2}\dfrac{a:S}{S:S}\label{eq:analyticalnutopt}\ ,
\end{eqnarray}
where the colon operator denotes double contraction. More specifically, for two Cartesian tensors $A$ and $B$ of rank two, their double contraction is given by $A:B\equiv A_{ij}B_{ij} = \text{tr}(AB^T) = {\rm tr}(BA^T)$ (Einstein summation convention implied on repeated indices), where $AB$ denotes the matrix product of $A$ and $B$. For the {\em special} case where both $A$ and $B$ are symmetric tensors (as in the current case), $A:B = {\rm tr}({AB}) = {\rm tr}({BA})$.

While optimal, the formulation in Eq.~(\ref{eq:analyticalnutopt}) has the disadvantage that $\nu^*_t$ can become negative (especially when the eigenframes of $a$ and $S$ are poorly aligned). When treated implicitly in Eq.~(\ref{eq:momentum}), a negative eddy viscosity can result in a combined negative effective viscosity, greatly destabilizing the numerical solution. Therefore, we propose the following formulation for the estimation of the optimal eddy viscosity:
\begin{eqnarray}
\nu^\dagger_t = \text{arg min}_{\nu_t\geq 0} \|a - (-2 \nu_t S)\|\ ,\label{eq:nnlsnut}
\end{eqnarray}
where $\nu^\dagger_t$  is the optimal eddy viscosity obtained using a non-negative least-squares approximation. The formulation in Eq.~(\ref{eq:nnlsnut}) guarantees that the injected eddy viscosity will not destabilize the iterative solution, at least in terms of reducing the diagonal dominance of the (assembled) coefficient matrix resulting from the discretization of the mean momentum transport equation. This formulation also allows an non-negativity constraint to be enforced on the output of a predictive model, further guaranteeing that any erroneous predictions at testing time do not destabilize the iterative solution.

In summary, the present optimal eddy viscosity formulation ($\nu^\dagger_t$) is the non-negative least-squares fit for $a\approx -2 \nu_t S$. This formulation harmonizes the conventional eddy viscosity approximation used in turbulence modelling, which is in terms of the anisotropy tensor, and promotes iterative stability by guaranteeing a non-negative effective viscosity.

\subsection{Non-linear part of the Reynolds stress tensor}\label{sec:aperp}
The formulation in Eq.~(\ref{eq:nnlsnut}) optimizes the \textit{linear} eddy viscosity approximation for $a$. An important question that arises is: how accurate is this linear approximation itself?

\begin{figure}
\includegraphics[]{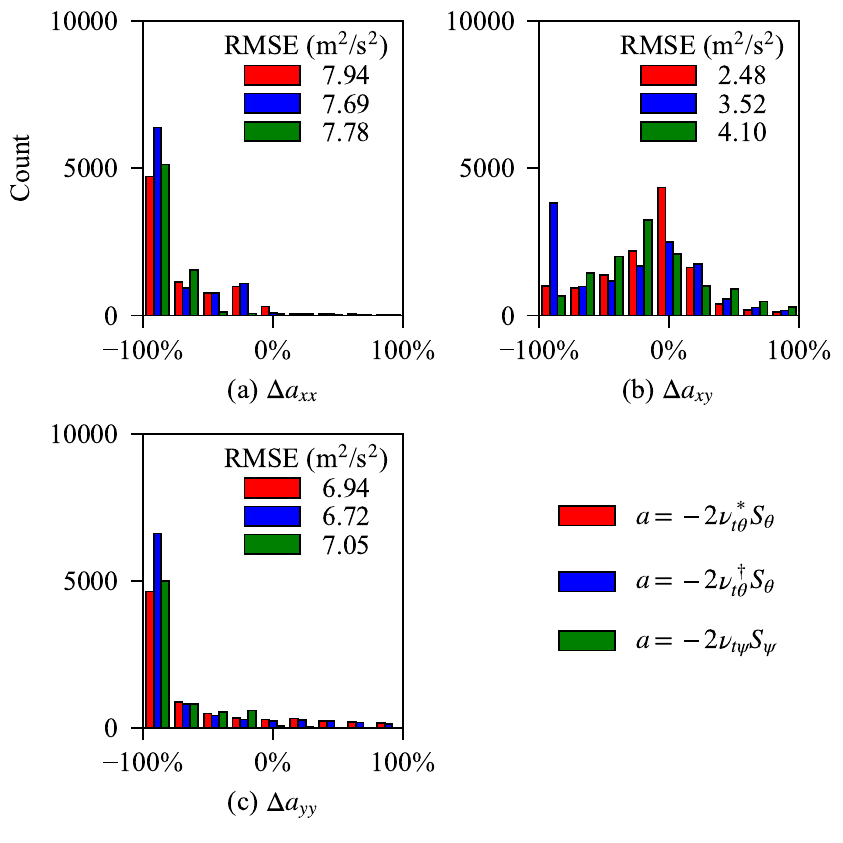}
\caption{\label{fig:Eddy_viscosity_error_a_theta_histograms} Error distribution in each anisotropy component after invoking the linear eddy viscosity approximation $a=-2\nu_t S$. Relative error is calculated by $\Delta a_{ij} = (a_{ij}-a_{ij\theta})/a_{ij\theta}$. $\nu^*_{t\theta}$ is the optimal eddy viscosity calculated using Eq.~(\ref{eq:analyticalnutopt}), and $\nu^\dagger_{t\theta}$ is the optimal eddy viscosity calculated using Eq.~(\ref{eq:nnlsnut}). The subscript $\theta$ indicates a quantity from DNS, and the subscript $\psi$ indicates a quantity from a RANS simulation using the $\phi$-$f$ model (Section~\ref{sec:phif}). The RMSE for each component of $a$ is shown in each plot.}
\end{figure}

To investigate this question, we consider a case of turbulent flow over periodic hills, simulated using both RANS~\cite{McConkey2021b} and direct numerical simulation (DNS)~\cite{Xiao2020}. Specifically, the $\alpha=1.2$ case is considered, where $\alpha$ is the slope steepness factor. Figure~\ref{fig:Eddy_viscosity_error_a_theta_histograms} shows the distributions of the error in $a$ after applying a linear eddy viscosity approximation for this canonical flow. Since this flow is two-dimensional, there are only three independent components of $a$: namely, $a_{xx}$, $a_{xy}$, and $a_{yy}$ (owing to the fact that $a$ is traceless). In the present work, the subscript $\theta$ is used for quantities obtained from a high-resolution DNS simulation, and the subscript $\psi$ indicates a quantity obtained from a RANS closure model. Even though the linear eddy viscosity approximation has been optimized on high-fidelity data for $a=-2\nu^*_{t\theta}S_\theta$ and $a=-2\nu^\dagger_{t\theta}S_\theta$, errors on the order of $100\%$ are present in all components of $a$. In particular, the principal components of $a$ are severely under predicted, even with an optimized eddy viscosity. Compared to the base RANS simulation, an optimal eddy viscosity approach applied to high-fidelity DNS data does not result in substantial accuracy gain.

\begin{figure}
\includegraphics[]{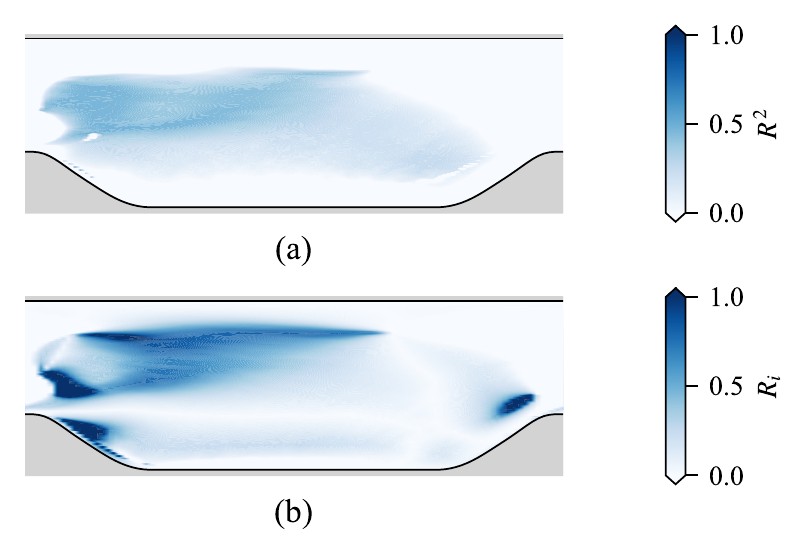}
\caption{\label{fig:R2_R_contours} Fit quality metrics after invoking the linear eddy viscosity approximation $a = -2 \nu^\dagger_{t\theta} S_\theta$: (a) $R^2$ value and (b) eddy viscosity model fit factor $R_i$ [Eq.~(\ref{eq:thompson})].}
\end{figure}

Figure~\ref{fig:R2_R_contours} shows two fit quality metrics for the approximation $a\approx -2 \nu^\dagger_t S$. In Fig.~\ref{fig:R2_R_contours}(a), the local $R^2$ value for the non-negative least squares fit $a_{ij\theta} \approx -2 \nu^\dagger_{t\theta} S_{ij\theta}$ is shown. While the $R^2$ values are higher near the bulk flow above the separated region, the $R^2$ values are generally poor near the walls, in the separated region, and during reattachment. Figure~\ref{fig:R2_R_contours}(b) shows the eddy viscosity fit quality metric proposed by Thompson {\em et al.}~\cite{Thompson2010} which is expressed as
\begin{eqnarray}\label{eq:thompson}
R_i = 1 - \frac{2}{\pi}\cos^{-1}\left( \sqrt{\frac{{\rm tr}(4\nu_{t\theta}^{\dagger 2} S^2_\theta)}{{\rm tr}(a^2_\theta)}}\right)\ .
\end{eqnarray}
This second fit quality metric generally agrees with the $R^2$ value in the bulk flow, in supporting a reasonable quality of the linear eddy viscosity approximation. However, for the separation over the left hill, and the acceleration over the right hill, higher values of the $R_i$ metric suggest a better quality in the optimal linear eddy viscosity fit. Nevertheless, visualizing both of these quality metrics demonstrates that there is substantial room for improvement in terms of representing $a$ using only a linear eddy viscosity approximation for this separated flow.


\begin{figure*}
\includegraphics{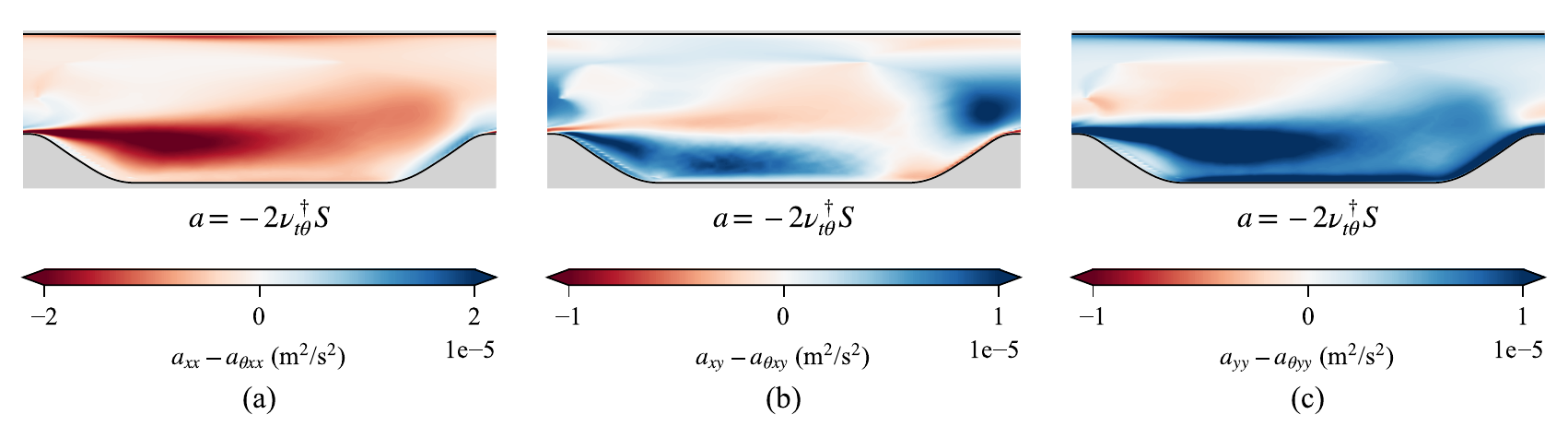}
\caption{\label{fig:a_error_contours} \textit{A priori} contours of error in each component of $a$ after invoking the linear eddy viscosity approximation $a=-2\nu^\dagger_{t\theta}S_\theta$.}
\end{figure*}

The errors in each component of $a$ arising from the linear eddy viscosity approximation are shown in Fig.~\ref{fig:a_error_contours}. The linear eddy viscosity approximation fails significantly for all components of $a$ during separation of the flow, a well-known deficiency~\cite{Pope2000}. Noticeable errors also exists in $a_{yy}$ during reattachment of the flow along the bottom wall. Though Thompson {\em et al.}'s $R_i$ value suggests a good approximation for $a$ as the flow accelerates over the right hill, the largest errors in the shear component $a_{xy}$ occur here.

\begin{figure}
\includegraphics[]{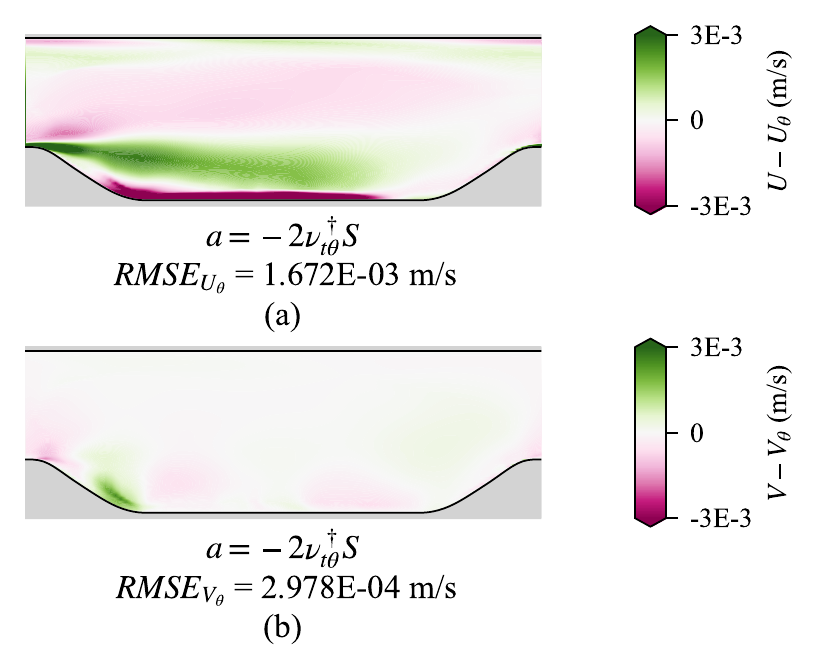}
\caption{\label{fig:Conditioning_error_linear} A poteriori contours of error in the velocity vector components after injecting the optimal linear eddy viscosity into the RANS equations, without any non-linear terms: (a) error in the $x$-component and (b) error in the $y$-component.}
\end{figure}

While the linear eddy viscosity approximation, even after optimization, was found to be deficient in approximating $a$, it was of interest to see whether these errors would affect the accuracy of the mean velocity field. Figure~\ref{fig:Conditioning_error_linear} shows the errors in the mean field components after injecting the optimal eddy viscosity implicitly into the RANS equation, and allowing the solution to converge around the fixed $\nu^\dagger_{t\theta}$. Figure~\ref{fig:Conditioning_error_linear}(a) shows that significant errors occur during separation and reattachment along the bottom wall in the streamwise velocity component $U$. 

Even though the solution in Fig.~\ref{fig:Conditioning_error_linear} was produced by injecting a high-fidelity optimal eddy viscosity into the RANS equations, there are significant errors in the mean flow field. Furthermore, the analysis in Section \ref{sec:conditioning} shows that conditioning errors with this optimal eddy viscosity formulation are minimal. If the target of the machine learning model is only the optimal eddy viscosity, mean field accuracy improvements are not guaranteed. In fact, the error fields shown in Fig.~\ref{fig:Conditioning_error_linear} represent the ``upper limit'' that could be achieved using a linear-component-only data-driven closure based on the optimal eddy viscosity as formulated in the present work.

Given the deficiencies of the linear eddy viscosity approximation, we propose an additional non-linear term in the anisotropy representation; namely, $a^\perp$ which is obtained from the following proportional/orthogonal tensor decomposition of $a$ :
\begin{eqnarray}
a = -2 \nu^\dagger_t S + a^\perp.\label{eq:a_decomp}
\end{eqnarray}
In Eq.~(\ref{eq:a_decomp}), $a$ has been decomposed into a linear component ($-2 \nu^\dagger_t S$) and a non-linear component ($a^\perp$). After calculating the optimal eddy viscosity using Eq.~(\ref{eq:nnlsnut}), the non-linear portion of the anisotropy tensor can be calculated using $a^\perp = a - (-2\nu^\dagger_t S)$.

\subsection{Conditioning}\label{sec:conditioning}

A subject that is of immense importance for data-driven turbulence closure modelling is the issue of conditioning within the RANS equations. When the RANS equations are ill-conditioned, small errors in the closure term can be amplified, resulting in large errors in the mean flow field. This issue affects data-driven models because an error in the model prediction of the closure term is almost always expected. Even with an error-free closure term, an accurate converged mean field cannot be guaranteed~\cite{Brener2021}. 

\begin{figure}
\includegraphics[]{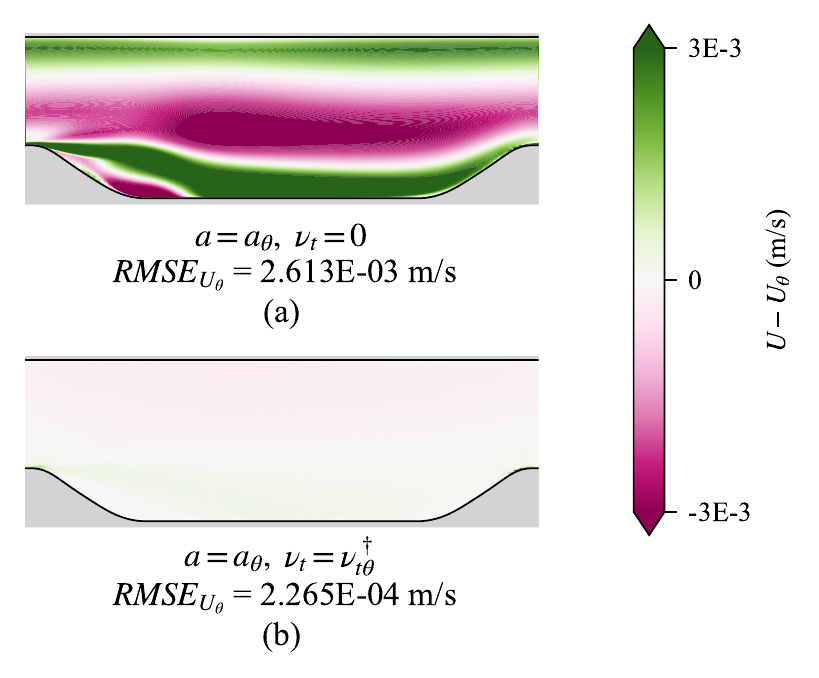}
\caption{\label{fig:conditioningU} \textit{A posteriori} contours of error in the $x$-velocity component after injecting the Reynolds stress anisotropy tensor from DNS into the RANS equations in two different ways: (a) fully explicit injection of $a_\theta$ and (b) the injection framework used in the present work.}
\end{figure}

\begin{figure}
\includegraphics[]{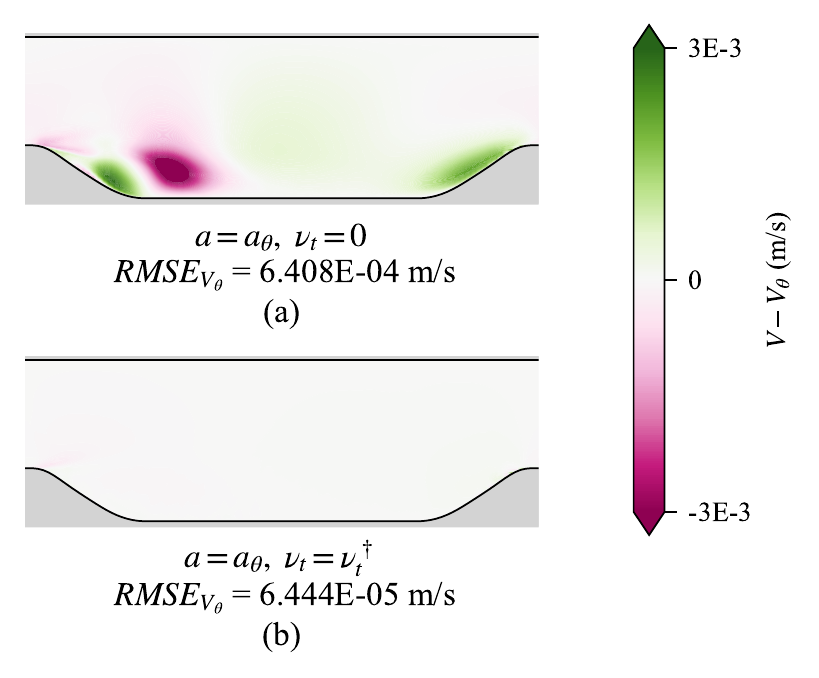}
\caption{\label{fig:conditioningV} \textit{A posteriori} contours of error in the $y$-velocity component after injecting the Reynolds stress anisotropy tensor from DNS into the RANS equations in two different ways: (a) fully explicit injection of $a_\theta$ and (b) the injection framework used in the present work.}
\end{figure}

To demonstrate the importance of the conditioning problem, a conditioning analysis similar to that performed by Brener {\em et al.}~\cite{Brener2021} was conducted. Figures~\ref{fig:conditioningU} and \ref{fig:conditioningV} show the errors in the mean velocity field components for two injection experiments. The flow considered here is a turbulent flow at $Re_H=5600$ over periodic hills\cite{Xiao2020}. In both Figs~\ref{fig:conditioningU}(a) and (b), the RANS equations have been closed using a highly accurate $a$ from a DNS simulation undertaken by Xiao {\em et al.}~\cite{Xiao2020}. In Figures~\ref{fig:conditioningU}(a) and \ref{fig:conditioningV}(a), $a_\theta$ has been injected as an explicit source term into the RANS equations. In Figures~\ref{fig:conditioningU}(b) and \ref{fig:conditioningV}(b), the decomposition proposed in the present work has been injected into the RANS equations in accordance to the following scheme: $\nu^\dagger_{t\theta}$ is treated implicitly and $a^\perp_\theta$ is treated as an explicit source term in the discretized mean momentum equations.

Despite both systems being closed by the same quantity arising from a high-fidelity simulation, the solution in Figs~\ref{fig:conditioningU}(a) and \ref{fig:conditioningV}(a) has a root mean square error (RMSE) that is an order of magnitude larger than the solution in Figs~\ref{fig:conditioningU}(b) and \ref{fig:conditioningV}(b). It should be noted that additional differences between the two solutions may arise from the need to use first-order schemes to stabilize the solution for the fully explicit propagation.

The low error observed after injecting the full anisotropy representation [Eq.~(\ref{eq:a_decomp})] from a DNS simulation into the RANS equation highlights the merits of the present approach. As discussed in Section~\ref{sec:deeplearning}, since the quantities $\nu^\dagger_{t\theta}$ and $a^\perp_\theta$ are the training labels, injecting these quantities into the RANS equations provides the ``upper limit'' that the data-driven model could achieve. For this reason, we recommend that data-driven closure frameworks be evaluated in terms of the conditioning errors that arise even after a perfect model prediction ($\tilde a=a_\theta$).



\subsection{Injection procedure}\label{sec:injection}

\begin{figure}
    \centering
    \includegraphics{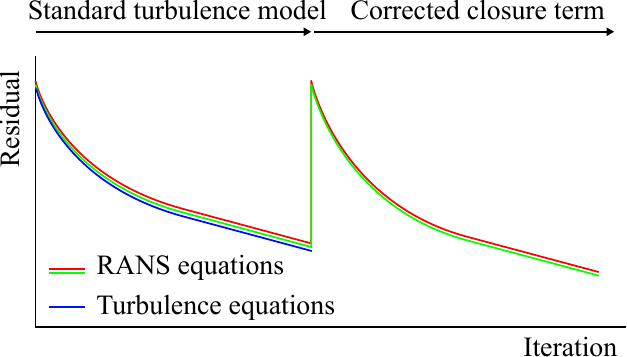}
    \caption{"Qualitative" residual plot for the machine learning corrective framework used in the present study. After injection, no further updates to the turbulence equations (e.g., the $k$ and $\varepsilon$ transport equations) are required.}
    \label{fig:injection}
\end{figure}

We use an open-loop, data-driven framework referred to by Ho and West\cite{Ho2021} as a "one-time correction" model. To this purpose, our framework involves making a fixed correction to the closure term, and then allowing the mean field to converge around this fixed correction. A qualitative description of the injection process in terms of the commonly used residual plot is shown in Figure~\ref{fig:injection}. The correction is estimated using a converged RANS simulation obtained from a base turbulence model as an input. This framework is in contrast to an iterative framework, where the closure term correction is updated repeatedly as the velocity field evolves. For example, the iterative framework used by Liu {\em et al.}~\cite{Liu2021} calls the machine learning model during each iteration. While open-loop frameworks are limited to steady flows, the design and use of iterative correction frameworks remains an open question. For example, the converged solution from an iterative framework can depend on the initial flow field~\cite{Ho2021}.

Motivated by the conditioning analysis of Brener {\em et al.}~\cite{Brener2021}, and our own analysis shown in Figs~\ref{fig:conditioningU} and \ref{fig:conditioningV}, we make two closure corrections in the present work: namely, one using $\nu^\dagger_t$ and another using $a^\perp$. We correct the eddy viscosity to be the optimal eddy viscosity, formulated in Eq.~(\ref{eq:nnlsnut}), and treat it implicitly in the momentum equation through an effective viscosity. Having established the importance of including the non-linear part of the anisotropy tensor (Section~\ref{sec:aperp}), we also include the divergence of $a^\perp$ as a source term in the momentum equation. 

After the correction, the RANS equations are solved iteratively until the mean fields converge around the fixed correction terms. The goal is that this converged solution will be more accurate than the solution produced by the base turbulence model. No update is required for the turbulent transport equations, as the eddy viscosity is directly corrected in our framework.

\subsection{The $\phi$-$f$ model}\label{sec:phif}
In the present work, we use the $\phi$-$f$ model implemented in OpenFOAM v2006 as the base RANS model~\cite{Laurence2005,Openfoam}. The $\phi$-$f$ model is a reformulated version of the $v^2$-$f$ model, where $\phi$ is the ratio $v^2/k$, and $f$ is a damping scalar. By adding a carefully constructed transport equation for the streamline-normal Reynolds stress component ($v^2$), this model accounts for the wall-blocking effects on the Reynolds stresses. The construction of the $v^2$ equation includes the effects of pressure strain on the streamline-normal Reynolds stress, a key contributor to wall-blocking effects. These wall-blocking effects are observed as unequal scaling (anisotropy) of the Reynolds stress components, which are not captured in a two-equation model. Often, the use of wall damping functions is required in two-equation models to correctly capture near-wall scaling. Laurence {\em et al.}'s~\cite{Laurence2005} $\phi$-$f$ model is a reformulated version of a $v^2$-$f$ model variant due to Lien and Kalitzin~\cite{Lien2001}, which is, in turn, an improved version of Durbin's~\cite{Durbin1991} original $v^2$-$f$ model.  Reformulating the $v^2$ and $f$ equation in terms of $\phi$ results in a more robust model. For example, this model has reduced stiffness in terms of the near-wall damping singularity \cite{Laurence2005}. The $\phi$-$f$ model estimates the eddy viscosity as
\begin{equation}
    \nu_t = C_\mu \phi k T \ ,
\end{equation}
where $C_\mu$ is a model coefficient, and $T$ is the turbulent time scale.

Details on the OpenFOAM implementation of this model are given in the OpenFOAM documentation \cite{Openfoam}. The original $\phi$-$f$ model equations can be summarized as follows:
\begin{equation}
    \frac{\partial k}{\partial t} + U_i \frac{\partial k}{\partial x_i} = P - \varepsilon + \frac{\partial}{\partial x_j}\left[ \left(\nu + \frac{\nu_t}{\sigma_k}\right) \frac{\partial k}{\partial x_j}\right]\ ;\label{eq:k}
\end{equation}
\begin{equation}
    \frac{\partial \varepsilon}{\partial t} + U_i \frac{\partial \varepsilon}{\partial x_i} = \frac{C_{\varepsilon_1} P}{T} - \frac{C_{\varepsilon_2}\varepsilon}{T} + \frac{\partial}{\partial x_j}\left[ \left(\nu + \frac{\nu_t}{\sigma_\varepsilon}\right) \frac{\partial \varepsilon}{\partial x_j}\right]\ ;\label{eq:epsilon}
\end{equation}
\begin{eqnarray}
    \frac{\partial \phi}{\partial t} + U_i \frac{\partial \phi}{\partial x_i} = f - P \frac{\phi}{k}+\frac{2\nu_t}{k\sigma_k}&&\frac{\partial \phi}{\partial x_j}\frac{\partial k}{\partial x_j} \nonumber\\ && + \frac{\partial}{\partial x_j}\left[ \left(\frac{\nu_t}{\sigma_k}\right) \frac{\partial \phi}{\partial x_j}\right]\ ;\label{eq:phi}
\end{eqnarray}
and,
\begin{eqnarray}
L^2 \frac{\partial^2 f}{\partial x_j} - f = \dfrac{1}{T}(C_{f_1} &&-1) \left[\phi - \dfrac{2}{3}\right] \nonumber\\ &&- C_{f_2} \dfrac{P}{k} - 2 \dfrac{\nu}{k}\dfrac{\partial \phi}{\partial x_j}\dfrac{\partial k}{\partial x_j} - \nu \dfrac{\partial^2 \phi}{\partial x_j}\ .\label{eq:f}
\end{eqnarray}
Equations~(\ref{eq:k}), (\ref{eq:epsilon}), and (\ref{eq:phi}) are model transport equations for $k$, $\varepsilon$, and $\phi$, respectively. Written in index notation, $U_i = \vec{U}$. Equation~(\ref{eq:f}) is an elliptic relaxation equation for $f$, which is a scalar predicting near-wall damping effects.

The turbulent time scale $T$ and length scale $L$ are given, respectively, by
\begin{eqnarray}
    T = \text{max}\left(\frac{k}{\varepsilon},C_T\sqrt{\frac{\nu}{\varepsilon}}\right)\label{eq:T} \ ,\\
    L = C_L\text{max}\left(\frac{k^{3/2}}{\varepsilon}, C_\eta \left(\frac{\nu^3}{\varepsilon}\right)^{1/4}\right) \ .\label{eq:L}
\end{eqnarray}
Finally, the model constants assume the values summarized 
as follows: $C_\mu = 0.22$; $C_{\varepsilon_1}=1.4(1.0+0.05 \sqrt{1.0/\phi})$; $ C_{\varepsilon_2}=1.9$; $C_T = 6.0$; $C_L = 0.25$; $C_{f_1} = 1.4$; $ C_{f_2} = 0.3$; $C_\eta = 110.0$; $\sigma_k = 1.0$; and, $\sigma_\varepsilon = 1.3$.

\subsection{Numerical methods}
All simulations in the present work use OpenFOAM v2006. For the base RANS simulation, the PIMPLE algorithm was used to achieve a converged solution. Then, a modified PIMPLE solver, which accepts the corrected $\nu^\dagger_t$ and $a^\perp$, was used to inject these corrections into the RANS equations and iteratively solve for the mean fields. While all flows in the present work reach a steady state solution, the unsteady algorithm was used to stabilize the iterative solution. 

For discretizing the RANS equations, a second-order upwind scheme was used for the convective terms, and a second-order central difference scheme was used for the diffusion terms. For the convective terms in the base turbulence model transport equations, a first-order scheme was used. 

A wall-resolved mesh ($y^+ \leq 1$) was used for all simulations. Further details on the mesh, domain, and boundary conditions are provided in McConkey {\em et al.}'s~\cite{McConkey2021b} description of the machine learning dataset. Cyclic boundary conditions were used for all flow variables at the inlet and outlet. At the top and bottom walls, the boundary conditions were fixed-zero for velocity, and zero-gradient for pressure.

\section{Deep learning procedure for predicting the Reynolds stress tensor}\label{sec:deeplearning}

\subsection{Dataset}
\begin{figure}
\centering
\includegraphics[width=0.5\textwidth]{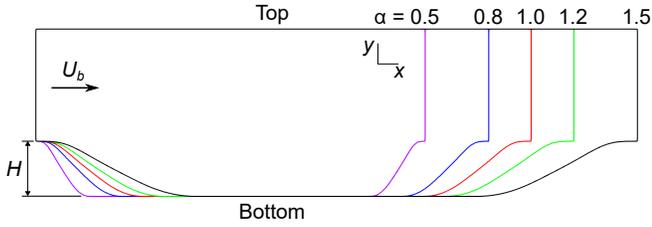}
\caption{\label{fig:geometry_phll} The geometry and flow variables used for the five periodic hills cases in the present work (Reproduced from McConkey {\em et al}.~\cite{McConkey2021b}).}
\end{figure}

To test the proposed decomposition and injection framework, a series of flow over periodic hills was chosen. This flow features separation over the left hill, reattachment along the bottom wall, and acceleration over the right hill. The dataset presented for data-driven turbulence modelling by McConkey {\em et al.}~\cite{McConkey2021b}, includes these cases, based on the DNS simulations of Xiao {\em et al.}~\cite{Xiao2020}. The periodic hills portion of the dataset consists of five cases, each with 14,751 data points, for a total of 73,755 data points. At each data point, the complete set of RANS input features and DNS fields are provided, so that the model can learn a mapping from the RANS features to the DNS labels. The RANS fields come from a converged solution using the $k$-$\varepsilon$-$\phi_t$-$f$ model \cite{Laurence2005} in OpenFOAM v2006. The five cases are generated by parametrically varying the hill steepness $\alpha$ and the overall length of the geometry. The five cases correspond to $\alpha = 0.5$, 0.8, 1.0, 1.2, and $1.5$, as shown in Fig.~\ref{fig:geometry_phll}.

\subsection{Model architecture}


The model consists of two independent neural networks: namely, EVNN, and apNN. The objective of EVNN is to predict the optimal eddy viscosity $\nu^\dagger_t$, whereas that of apNN is to predict the non-linear part of the anisotropy tensor, $a^\perp$. Both models are feed-forward, fully connected neural networks. The EVNN has one output neuron, and the apNN has three output neurons (one for each independent component of $a^\perp$).

Because $\nu^\dagger_t$ is non-negative, the stability of the EVNN predictions during training and injection can benefit by enforcing a non-negative prediction. For the EVNN, the activation function used for the output neuron was a simple exponential function $f(x) = e^x$, where $x$ is the input value to the layer. This activation function guarantees a non-negative eddy viscosity prediction by the EVNN, regardless of the input activation layer (viz., the choice of the exponential function as the activation function induces an inductive bias of the network output). This stability constraint is possible due to the formulation of $\nu^\dagger_t$, which guarantees a non-negative training label. For the apNN, the three output neurons utilize a linear activation function. For the hidden layers, the scaled exponential linear unit (SELU) activation function was used. Combined with the LeCun initialization of the network weights, a SELU-activated, fully connected, feed-forward neural network is self-normalizing~\cite{Klambauer2017}, eliminating the need for batch normalization layers~\cite{Geron}.

The neural networks were implemented using the high-level Keras application programming interface (API) in TensorFlow \cite{keras}.
The hyperparameters used for both networks were identical: namely, both networks have $L=14$ hidden layers, with $n=30$ neurons in each layer, resulting in a network architecture with an depth-to-width aspect ratio of $L/n \approx 0.467$. The EVNN has 13,141 trainable parameters (weights and biases) with a single output, and the apNN has \text{13,203} trainable parameters with three outputs. For traditional turbulence modelling, tuning this number of parameters is a nearly impossible task. However, for deep learning, tuning these parameters is relatively ``straightforward''---the number of learnable parameters is often orders of magnitude greater than those used in the EVNN and the apNN. Nevertheless, despite the large number of model parameters used in these types of overparameterized neural networks (where the number of model parameters far exceed the number of training data), there is mounting empirical (indeed, practical) evidence that these networks can be successfully trained to provide excellent predictions---perhaps supporting the notion that it is not the number of model parameters, but the representation learning provided by the subtle correlations in the intralayer and interlayer interactions of the neural units in the network that ultimately determines the predictive skill and generalizability of the model in deep learning.

\subsection{Input feature selection}
A wide range of input features have been used in previous data-driven closure frameworks. The input features for a data-driven closure are generally required to possess the same invariance properties as the underlying Navier-Stokes equations: reflection, rotation, and Galilean transformation invariance. Methods used to generate the input feature set vary from a purely heuristic selection, to a systematic generation of an integrity basis~\cite{Wu2018}. In the present work, we use a combination of these two methods. Three heuristic scalars are selected as input features given by
\begin{eqnarray}
q_1 &&\equiv \text{min}(2.0,\sqrt{k} y_\perp/(50\nu))\ , \nonumber\\
q_2 &&\equiv k/\varepsilon \|S\|\ , \nonumber\\
q_3 &&\equiv \|\tau\|/k\ .
\end{eqnarray}
Here, $y_\perp$ is the wall-normal distance. $q_1$ is the wall  distance based Reynolds number, $q_2$ is the ratio of the mean strain timescale to a turbulent time scale, and $q_3$ is the ratio of the total Reynolds stress magnitude to the diagonal Reynolds stresses. These Galilean invariant scalars were used previously by Kaandorp and Dwight~\cite{Kaandorp2020}.

In addition to these scalars, we modify the procedure presented by Wu {\em et al.}~\cite{Wu2018} to significantly augment the input feature set. In Wu {\em et al.}'s procedure, four gradient tensors were selected: $S$, $R$,$\nabla k$, and $\nabla p$. The TKE and pressure gradient vectors were converted to second-order tensors, by casting their components into an anti-symmetric matrix. Then, the method outlined by Spencer and Rivlin~\cite{Spencer1962} was used to generate an integrity basis for the space spanned by these four second-order tensors. This integrity basis consists of 47 tensors. Finally, a scalar input feature was extract from each tensor, by taking the first invariant (viz., the trace) of each integrity basis tensor.

While this procedure extracts a large number of input features from the flow, a number of issues arise that have received limited attention in the literature. Firstly, the terms related to $\nabla p$ in the integrity basis tend to be numerically unstable in our experience, possibly due to the normalization used by Wu {\em et al.}~\cite{Wu2018}. In their frameworks, Wang {\em et al.}~\cite{Wang2017} and Kaandorp and Dwight~\cite{Kaandorp2020} omitted these terms, reducing the number of invariants down to 16. Secondly, many of these invariants are zero for two-dimensional flows. These zero invariants arise from either a direct result of the $z$-direction gradients being zero, or from the incompressibility of the flow. Lastly, this procedure only uses the first invariant $I_1$ (or, trace) of each integrity basis tensor---more information can be extracted from this basis by including further invariants in the input feature set. We have presented an analysis of Wu {\em et al.}'s~\cite{Wu2018} integrity basis in Appendix \ref{ap:invariants}, showing expressions for each of the basis tensor invariants for two-dimensional flows.

To circumvent the issue of unstable terms related to $\nabla p$ being unusable as input features, we replace $\nabla p$ in Wu {\em et al.}'s basis with $\nabla v^2$. This replacement is possible since our base turbulence model is the $\phi$-$f$ model (Section~\ref{sec:phif}). We found that terms including $\nabla v^2$ were much more stable than the terms including $\nabla p$, and therefore more features could be included. It is noted that $\nabla v^2$ provides additional information compared to $\nabla k$, because $v^2$ measures the degree of wall-blocking effects and anisotropy in the flow. Additionally, we augment the 47 original first invariants by also taking the second invariant $I_2$ of a tensor $A$ ($I_2 = \tfrac{1}{2}[(\text{tr}(A))^2 - \text{tr}(A^2)]$). For a three-dimensional flow, this effectively doubles the number of basis tensor invariants that can be used as input features. However, as shown in Appendix~\ref{ap:invariants}, many of these invariants are zero for two-dimensional flows. Nevertheless, after eliminating the zero-valued invariants, a set of 29 invariants remained as suitable input features. Table~\ref{tbl:features} summarizes the input features used.

\begin{table}[]
\caption{Input features for EVNN and apNN.}\label{tbl:features}
\begin{tabular}{cccc}
\hline
Number & Input feature & Expression                           & Transformation      \\ \hline
1      & $I_1(B_1)$    & $I_1(S^2)$                           & $\text{log}(|x|+1)$   \\
2      & $I_1(B_3)$    & $I_1(R^2)$                           & $\text{log}(|x|+1)$ \\
3      & $I_1(B_4)$    & $I_1(A_{v2}^2)$                      & $\sqrt[3]{x}$       \\
4      & $I_1(B_5)$    & $I_1(A_k^2)$                         & $\text{log}(|x|+1)$ \\
5      & $I_1(B_7)$    & $I_1(R^2S^2)$                        & $\text{log}(|x|+1)$ \\
6      & $I_1(B_9)$    & $I_1(A_{v2}^2 S)$                    & $\sqrt[3]{x}$       \\
7      & $I_1(B_{10})$ & $I_1(A_{v2}^2 S^2)$                  & $\text{log}(|x|+1)$ \\
8      & $I_1(B_{12})$ & $I_1(A_{k}^2 S)$                     & $\sqrt[3]{x}$       \\
9      & $I_1(B_{13})$ & $I_1(A_{k}^2 S^2)$                   & $\text{log}(|x|+1)$ \\
10     & $I_1(B_{16})$ & $I_1(A_{v2}A_k)$                     & $\text{log}(|x|+1)$ \\
11     & $I_1(B_{21})$ & $I_1(A_{v2}RS)$                      & $\text{log}(|x|+1)$ \\
12     & $I_1(B_{25})$ & $I_1(A_{v2}^2SRS^2)$                 & $\text{log}(|x|+1)$ \\
13     & $I_1(B_{29})$ & $I_1(A_{k}^2RS)$                     & $\text{log}(|x|+1)$ \\
14     & $I_1(B_{33})$ & $I_1(A_{k}SRS^2)$                    & $\text{log}(|x|+1)$ \\
15     & $I_1(B_{34})$ & $I_1(A_{v2}A_kS)$                    & $\sqrt[3]{x}$       \\
16     & $I_1(B_{35})$ & $I_1(A_{v2}A_k S^2)$                 & $\text{log}(|x|+1)$ \\
17     & $I_1(B_{42})$ & $I_1(RA_{v2}A_k)$                    & $\sqrt[3]{x}$       \\
18     & $I_1(B_{43})$ & $I_1(RA_{v2}A_kS)$                   & $\sqrt[3]{x}$       \\
19     & $I_1(B_{45})$ & $I_1(RA_{v2}A_kS^2)$                 & $\sqrt[3]{x}$       \\
20     & $I_1(B_{46})$ & $I_1(RA_kA_{v2}S^2)$                 & $\sqrt[3]{x}$       \\
21     & $I_2(B_1)$    & $I_2(S^2)$                           & $\text{log}(|x|+1)$   \\
22     & $I_2(B_2)$    & $I_2(S^3)$                           & $\text{log}(|x|+1)$ \\
23     & $I_2(B_3)$    & $I_2(R^2)$                           & $\text{log}(|x|+1)$   \\
24     & $I_2(B_4)$    & $I_2(A_{v2}^2)$                      & $\text{log}(|x|+1)$   \\
25     & $I_2(B_5)$    & $I_2(A_{k}^2)$                       & $\text{log}(|x|+1)$   \\
26     & $I_2(B_6)$    & $I_2(R^2S)$                          & $\text{log}(|x|+1)$ \\
27     & $I_2(B_7)$    & $I_2(R^2S^2)$                        & $\text{log}(|x|+1)$   \\
28     & $I_2(B_8)$    & $I_2(R^2SRS^2)$                      & $\text{log}(|x|+1)$ \\
29     & $I_2(B_{16})$ & $I_2(A_{v2}A_k)$                     & $\text{log}(|x|+1)$   \\
30     & $q_1$         & $\text{min}(2.0,\sqrt{k} y_\perp/(50\nu))$ & ---                   \\
31     & $q_2$         & $k/\varepsilon \|S\|$                & ---                   \\
32     & $q_3$         & $\|\tau\|/k$                         & ---                   \\
33     & $\phi$        & $\phi$                               & ---                   \\ \hline
\end{tabular}
\end{table}
This input feature set represents one of the richest feature sets used in a data-driven turbulence closure model. While many studies relax the condition of Galilean invariance (most commonly, by including the turbulence intensity), all of our input features are Galilean invariant. Section~\ref{sec:interpret} analyzes the relative importance of each of these features to the model predictions.

\subsection{Label calculation}
Each neural network has a different set of training labels, based on the desired output. For the EVNN, the goal is to predict the optimal eddy viscosity $\nu^\dagger_t$. As shown in Section~\ref{sec:conditioning}, injecting the optimal eddy viscosity from DNS produced a well-conditioned closure. Therefore, the label field for the EVNN consists of the $\nu^\dagger_{t\theta}$ field. To calculate this field, scikit-learn \cite{scikit-learn} was used to perform the non-negative least-squares (NNLS) regression fit for $a_\theta = -2\nu_tS_\theta$ at each cell in the dataset.

The labels for the apNN consist of the three independent components of the non-linear part of the Reynolds stress tensor. Therefore, the three component labels were extracted from the tensor $a^\perp_\theta = a_\theta + 2 \nu^\dagger_{t\theta}S_\theta$, where $\nu^\dagger_{t\theta}$ are the labels for the EVNN. At injection time, $a^\perp_{zz}$ can be calculated as $a^\perp_{zz}=-a^\perp_{xx} - a^\perp_{yy}$, because $a^\perp$ is a traceless tensor.

\subsection{Pre-processing}
\label{sec:preprocessing}

After calculating the input feature and label sets, the five periodic hills cases were divided into a training set, a validation set, and a testing set. The training process consists of updating the model weights and biases using an optimization algorithm. The model makes predictions using the training set features and evaluates the predictions using the training set labels. The validation and testing sets are used similarly, but at different times in the process. The validation set is used to evaluate the predictive performance of the neural network throughout the training process in order to assess when the model has converged. The validation set is not used to update the weights and biases of the network. Instead, it helps identify overfitting during the training process. If the training process continues for too long, the model will begin to memorize noise and other undesired details in the training set. Repeated evaluation of the neural network predictive performance using the validation set will allow one to assess the generalization performance of the model during training. The testing set is used to evaluate the final, trained model on an entirely new set of data---it can be used to determine the model (or generalization) error and to evaluate the generalization capability of the fully-trained neural network model. More specifically, the generalization error is perhaps the primary quantitative measure of the success of the fully-trained neural network model providing, as such, the standard for how well the network is really approximating the underlying function for the Reynolds stresses.

The $\alpha=1.2$ case was selected as a test set. In terms of the parameter $\alpha$, this represents an interpolation test case, since the training and validation set contain $\alpha$ values both above and below 1.2. The training/validation set consists of 59,004 data points, for the remaining four cases. At this stage, the typical procedure is to shuffle and split all of the remaining data into a training and validation set. However, for training a data-driven turbulence closure neural network model, we propose a different method to better estimate the generalization performance during training. Our recommendation is that the validation set should also consist of distinct cases from the training set. For example, the validation set could be all data points for the $\alpha=1.0$ test case, and the training data would then consist of the $\alpha = 0.5$, 0.8, and 1.5 cases.

It is important to stress that the choice of validation set greatly influences the training process. This issue is normally solved by using $k$-fold cross-validation, where the training data is split into $k$ folds. Each of the folds is then used as a validation set, while the model is trained on the remaining folds. We apply 4-fold cross-validation, where the model is trained four times, each using a remaining case ($\alpha = 0.5$, 0.8, 1.0, 1.5) as the validation set. The model which exhibits the best generalization performance (defined as the difference between the training and validation error) is then selected for testing.

The input features and labels are often scaled and transformed in deep learning in order to eliminate scale differences between input features and to ensure well behaved weights and biases during training~\cite{Geron}. To this purpose, a transformation was selected for each of the input features and labels to ensure a more uniform distribution (histogram) of their values before they are used in training. These transformations are given in Table~\ref{tbl:features}. The selection of input feature transformations is highly heuristic, and depends on the data used for each study. However, we have generally applied a logarithm transform ($\hat{x} = \log (|x|+1)$) to data which is entirely positively- or negatively-valued, and a cube-root transform ($\hat{x} = x^{1/3}$) to data which contains both positive and negative values. For the majority of data containing RANS mean fields, the histograms tend to be highly skewed due to domain-specific factors. For example, the behavior of almost all fields in the near-wall region is completely different than the bulk flow region. It was found that transforming the components of $a^\perp$ was not necessary for sufficient model performance. After transforming the features (Table~\ref{tbl:features}), the values were scaled to lie in the unit interval [0,1] using scikit-learn's MinMaxScaler~\cite{scikit-learn}. The $\nu^\dagger_{t\theta}$ labels were transformed using $\hat{ x} = \text{log}(|x| +1)$, then scaled using the same method as the features. No transformation was necessary for the $a^\perp_{\theta}$ labels. Only scaling was applied to the $a^\perp_{\theta}$ label set.

\subsection{Training}
For the training process, the Nadam optimizer in Keras was selected~\cite{keras}. This optimizer combines the adaptive moment estimation (Adam) optimizer with Nesterov momentum to accelerate training. The learning rate used was $1\times 10^{-4}$. For the EVNN, the loss function was the mean-squared error, with L2 regularization used. The L2 regularization weight $\lambda$ was $1\times 10^{-5}$. L2 regularization adds an additional penalty term to the loss function to constrain the norm squared of the optimized model parameters (or, weights)---the neural network seeks the best compromise between minimizing the errors in the prediction of the output and constraining the magnitude of the neural network weights to be as small as possible (viz., shrinking the weights towards zero). In practice, this regularization is required for some deep learning scenarios. We did not apply any L2 regularization to the apNN loss function, as the generalization performance was found to be sufficient without any additional regularization term to constrain the weights.

\begin{figure}
\includegraphics[]{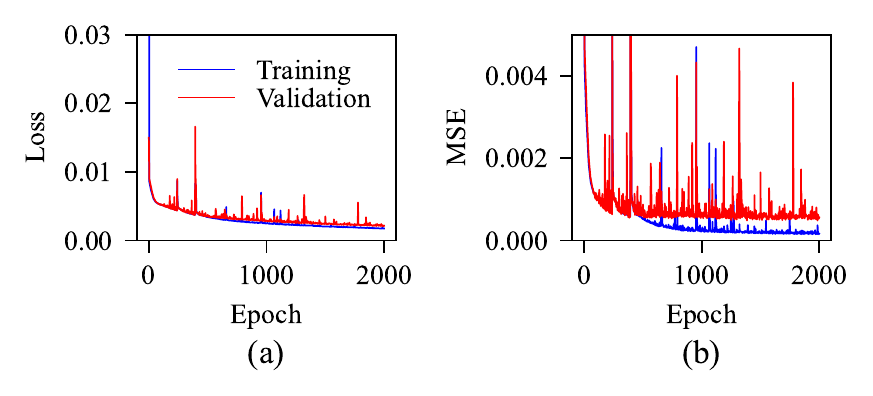}
\caption{\label{fig:loss_EVNN} Loss curves during training for the EVNN: (a) the total loss function consisting of the mean squared error (MSE) and the L2 regularization (penalty term) of the weights and (b) the MSE portion of the total loss function.}
\end{figure}

\begin{figure}
\includegraphics[]{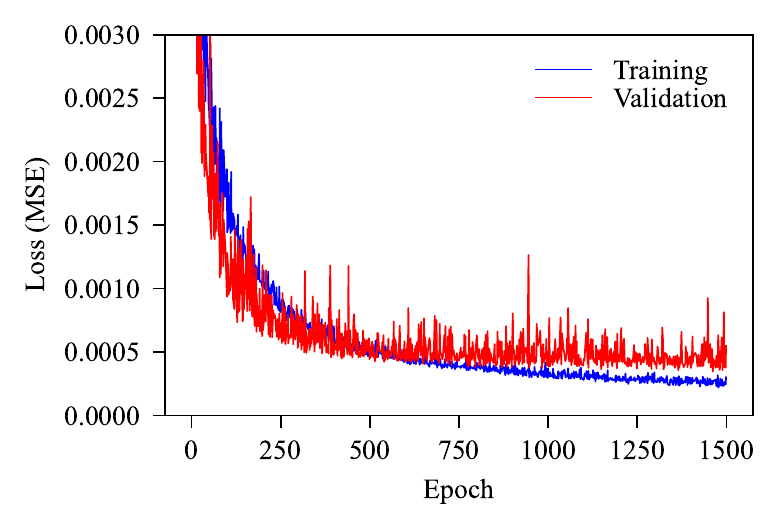}
\caption{\label{fig:loss_apNN} Loss curve during training for the apNN. The loss function for the apNN consists of the mean squared error (MSE) only (viz., no regularization term was added to the loss function).}
\end{figure}

The training process was completed for the four cross-validation folds, as described in Section~\ref{sec:preprocessing}. After completing the training procedure for each of these folds, the model which provided the best generalization performance was selected. The training loss curves for the selected EVNN and apNN are shown in Figs~\ref{fig:loss_EVNN} and \ref{fig:loss_apNN}, respectively. While the apNN loss curve remains relatively stable, the loss curve for the EVNN contained instabilities throughout the training process. Typically, these instabilities are a symptom of the learning rate being too large, but the learning rate was not found to be responsible for the occurrence of these jumps. This indicates that there may be significant discontinuities in the total loss surface consisting of the mean-squared error for prediction of $\nu_t^\dagger$ and the L2 regularization (penalty) term involving the weights. After approximately 1000 epochs, the mean-square error (MSE) part of the EVNN loss function remains relatively stable, indicating that no further accuracy is being gained in predicting $\nu^\dagger_t$. The remaining drop in the overall loss function comes from the training procedure optimizing the weights in the EVNN. It was found that the optimal number of epochs at which to stop the EVNN training was 2000, which allows an additional 1000 epochs for the optimizer to tune the network weights.

The loss curve for the apNN (cf.~Fig.~\ref{fig:loss_apNN}) is simply the MSE of the prediction, because no regularization term for the weights was included in the training process for this neural network. After about 500 epochs, the training loss is smaller than the validation loss, and the training loss continues to decrease monotonically after this point. By assessing the {\em a priori} and {\em a posteriori} performance for various selections of the number of epochs to use as the training endpoint (early stopping), it was found that a training endpoint of 1500 epochs yielded the best results for the apNN. 

\section{Results}\label{sec:results}
The final step in the deep learning procedure includes a test on a hold-out case---known also as the test case that is set aside and used to evaluate the neural network model only after training is complete. For the periodic hills training set, the $\alpha=1.2$ case was held out of the training. Testing the network on a hold-out case provides an assessment of the network's ability to generalize by predicting a previously unseen case. The results for the hold-out test case are discussed in the following sections. A tilde is used to denote a quantity predicted by the neural networks (e.g., $\tilde \nu^\dagger_t$ is a prediction of the optimal eddy viscosity provided by the neural network).

For the open-loop framework used here, there are two stages of evaluation: namely, {\em a priori} and {\em a posteriori}. In Section~\ref{sec:apriori}, the model predictions are evaluated before injecting into the modified PIMPLE solver ({\em a priori} assessment). In Section~\ref{sec:aposteriori}, the resulting mean flow fields and combined anisotropy tensor ($a$) are evaluated after the iterative solution has converged around the fixed predictions ({\em a posteriori} assessment). The framework used here treats the linear component of $a$ implicitly, so the combined prediction $\tilde a$ is not available in an {\em a priori} sense---it must be evaluated {\em a posteriori}.

\subsection{{\em A priori} assessment}\label{sec:apriori}
\begin{figure}
\includegraphics[]{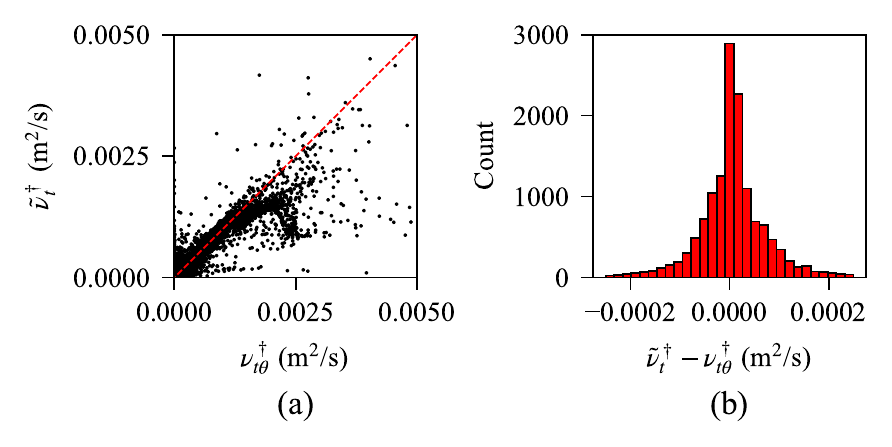}
\caption{\label{fig:nut_pred_hist} \textit{A priori} prediction accuracy for the EVNN on the $\alpha=1.2$ test case: (a) plot of the predicted value (ordinate) vs the ground-truth value obtained from DNS (abscissa) and (b) distribution of the $\nu^\dagger_t$ prediction errors. }
\end{figure}

Prior to injecting the model predictions into the RANS equations, the predictions on the $\alpha=1.2$ test case were analyzed for accuracy. The objective of the EVNN is to accurately predict $\nu^\dagger_t$. The mean-squared error is calculated by (for $N$ data points)
\begin{equation}
{\rm MSE} = \frac{1}{N}{\sum_{i=1}^N (\tilde \nu^\dagger_{ti} - \nu^\dagger_{t\theta i})^2}\ .
\end{equation}
For the $\alpha=1.2$ test case, the MSE in the prediction of $\nu^\dagger_t$ is $1.89\times 10^{-7}$ m$^4$~s$^{-2}$. Figure \ref{fig:nut_pred_hist} compares $\tilde \nu^\dagger_t$ to $\nu^\dagger_{t\theta}$. Figure~\ref{fig:nut_pred_hist}(a) shows that while the predictions generally fall along the (ideal) line $\tilde \nu^\dagger_t = \nu^\dagger_{t\theta}$ at smaller values of $\nu^\dagger_t$, some scatter in the predictions is observed at larger values. The error distribution for the prediction of $\nu^\dagger_t$ [Fig.~\ref{fig:nut_pred_hist}(b)] is roughly symmetric, indicating that there is not a significant over or under prediction tendency for the EVNN. The large prediction errors in Fig.~\ref{fig:nut_pred_hist} demonstrate the importance of model conditioning---these errors will always be present at model testing time.

\begin{figure}
\includegraphics[]{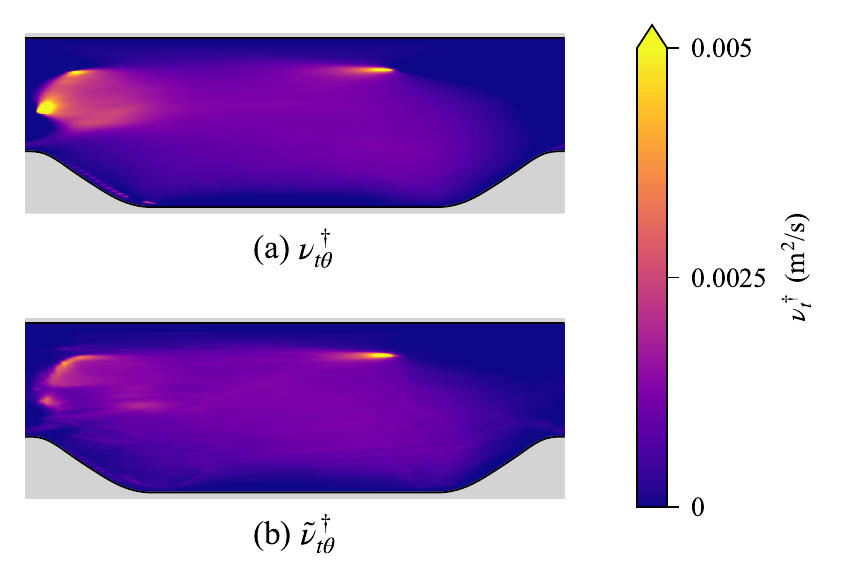}
\caption{\label{fig:nut_pred_contours} \textit{A priori} contours of $\nu^\dagger_t$ for the test case: (a) ground-truth value from DNS and (b) value predicted by the EVNN.}
\end{figure}

Figure~\ref{fig:nut_pred_contours} shows the spatial contours of $\tilde \nu^\dagger_t$ and $\nu^\dagger_{t\theta}$.
This figure shows that the EVNN is able to capture trends in the spatial variation of $\nu^\dagger_t$ well. The large values for $\nu^\dagger_t$ in the center of the domain are predicted well by the EVNN. In general, the neural network predictions match the ground truth (DNS data), with some minor discrepancies in the high $\nu^\dagger_t$ region above the first crest. While Fig.~\ref{fig:nut_pred_hist} presents a pessimistic view of the ability of the EVNN to accurately predict $\nu^\dagger_t$, the contours in Fig.~\ref{fig:nut_pred_contours} provide a more optimistic view.

\begin{figure}
\includegraphics[]{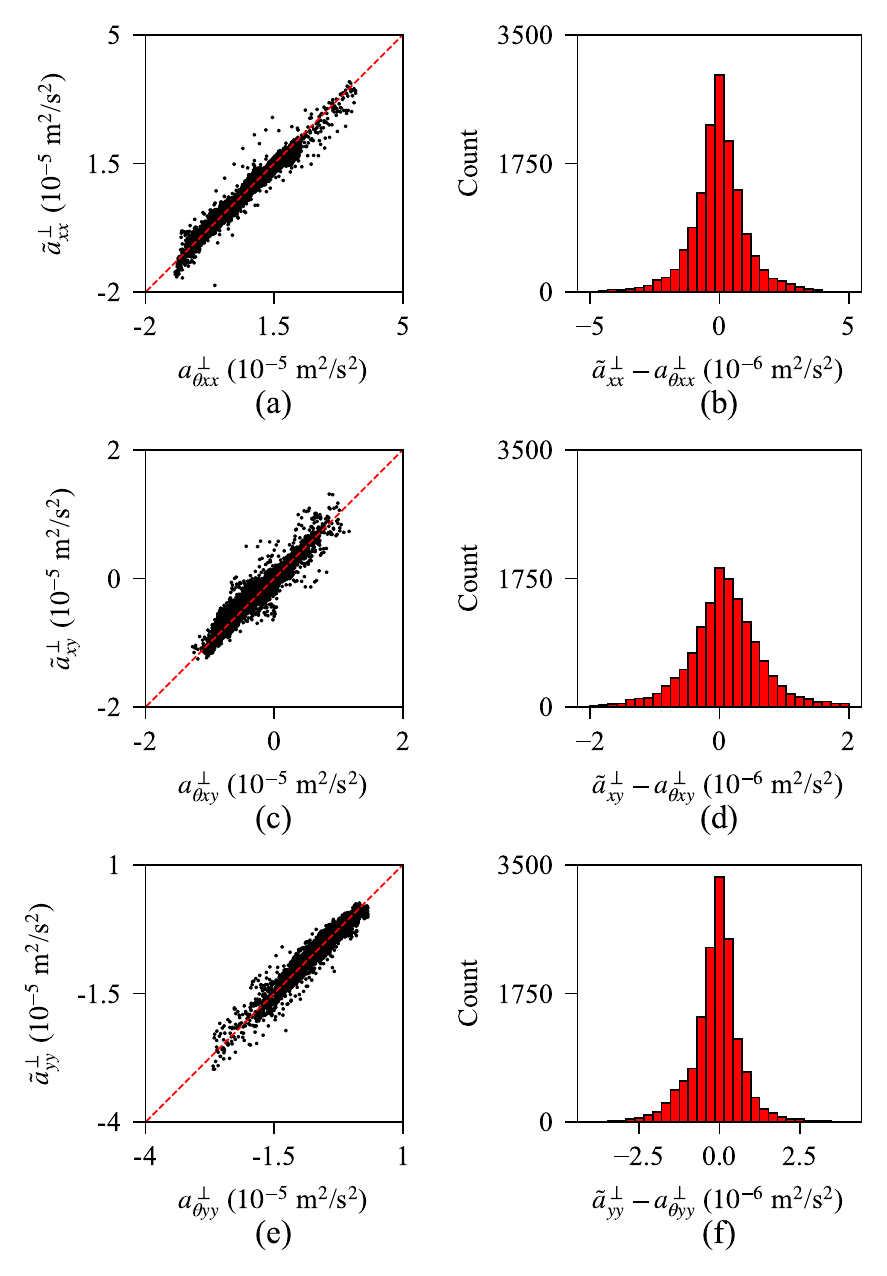}
\caption{\label{fig:ap_pred_hist} \textit{A priori} prediction accuracy for the apNN on the $\alpha=1.2$ test case. The prediction accuracy for each component of $a^\perp$ is shown: (a), (c), (e) plots of the predicted value (ordinate) vs the ground-truth value obtained from DNS (abscissa) and (b), (d), (f) distribution of the $a^\perp$ prediction errors. }
\end{figure}

\begin{figure*}
\includegraphics[]{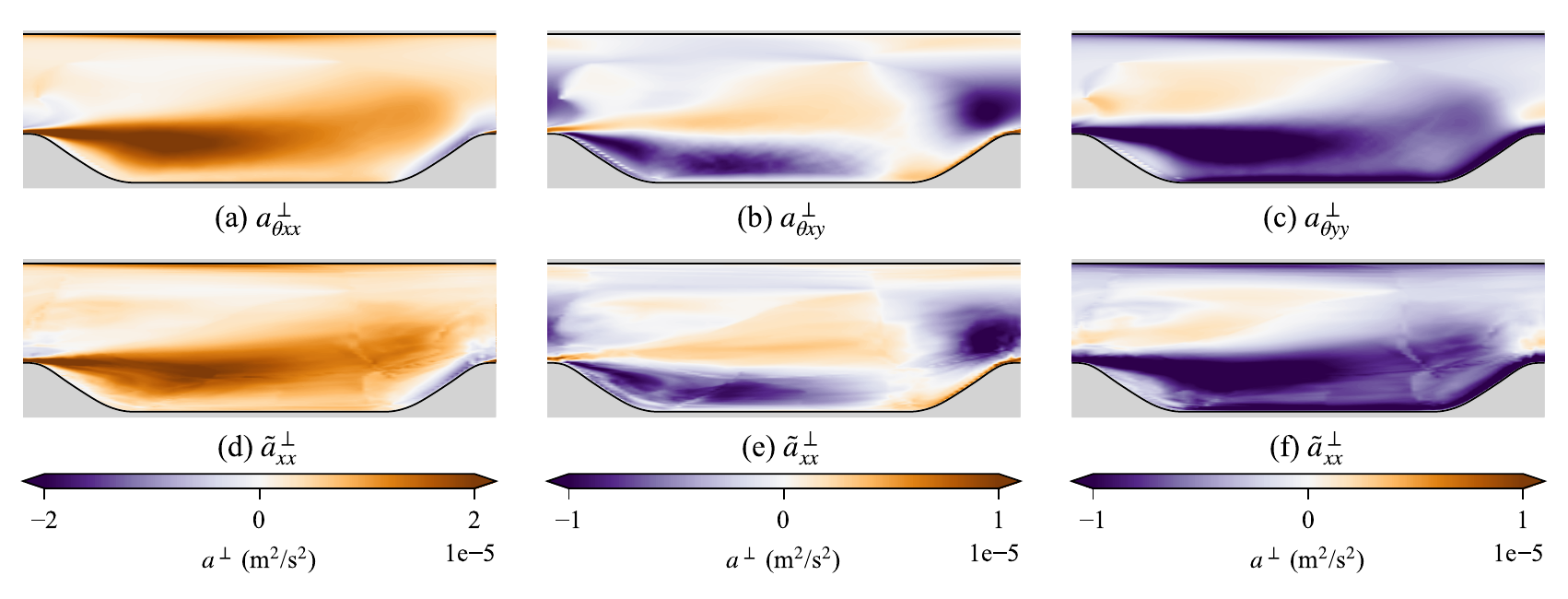}
\caption{\label{fig:ap_pred_contour} \textit{A priori} contours of each component of $a^\perp$ for the test case: (a), (b), (c) ground-truth value obtained from DNS and (d), (e), (f) value predicted by the apNN.}
\end{figure*}

Figures~\ref{fig:ap_pred_hist} and \ref{fig:ap_pred_contour} show that the apNN predicts the non-linear part of the anisotropy tensor well. The MSEs associated with the prediction of each component of $a^\perp$ are ${\rm MSE}_{a^\perp_{xx}}=2.31\times 10^{-12}$, MSE$_{a^\perp_{xy}}=9.40\times 10^{-13}$, and MSE$_{a^\perp_{yy}}=9.56\times 10^{-13}$ (in units of m$^4$~s$^{-4}$). Spatial trends in the magnitude of $a^\perp$ are accurately predicted by the apNN (cf.~Fig.~\ref{fig:ap_pred_contour}). The individual points for the various components generally lie all along the ideal line $\tilde{a}^\perp=a^\perp_\theta$ (shown as the diagonal dashed lines in Fig.~\ref{fig:ap_pred_hist}). A perusal of Fig.~\ref{fig:ap_pred_hist} shows that the diagonal (normal stress) components of $a^\perp$ are generally better predicted using apNN than the off-diagonal (shear stress) components. Indeed, the error distributions for the prediction of the normal stress components are more markedly peaked around zero (error) than that for the shear stress component [cf.~Figs.~\ref{fig:ap_pred_hist}(a) and (f) with Fig.~\ref{fig:ap_pred_hist}(d)].


Comparing the predictions by the EVNN (Figs~\ref{fig:nut_pred_hist} and \ref{fig:nut_pred_contours}) to the predictions by the apNN (Figs~\ref{fig:ap_pred_hist} and \ref{fig:ap_pred_contour}), we see that the apNN generally out-performs the EVNN in terms of predicting their respective targets. This outcome agrees with our experience training and testing several times with these models---$\nu^\dagger_t$ is harder to predict than 
$a^\perp$. This discrepancy indicates that the information content embodied in the RANS fields for the prediction of $a^\perp$ is larger than that for the prediction of $\nu^\dagger_{t\theta}$.

\subsection{{\em A posteriori} assessment}\label{sec:aposteriori}
The fields predicted by the EVNN and apNN were injected into the RANS equations. A similar procedure as Section \ref{sec:conditioning} was utilized, whereby a modified PIMPLE solver in OpenFOAM v2006 was employed to inject the predicted fields from the EVNN ($\tilde \nu_t^\dagger$) and apNN ($a^\perp$) into the RANS equations, and iteratively reach a converged solution for the mean velocity field. 

\begin{figure}
\includegraphics[]{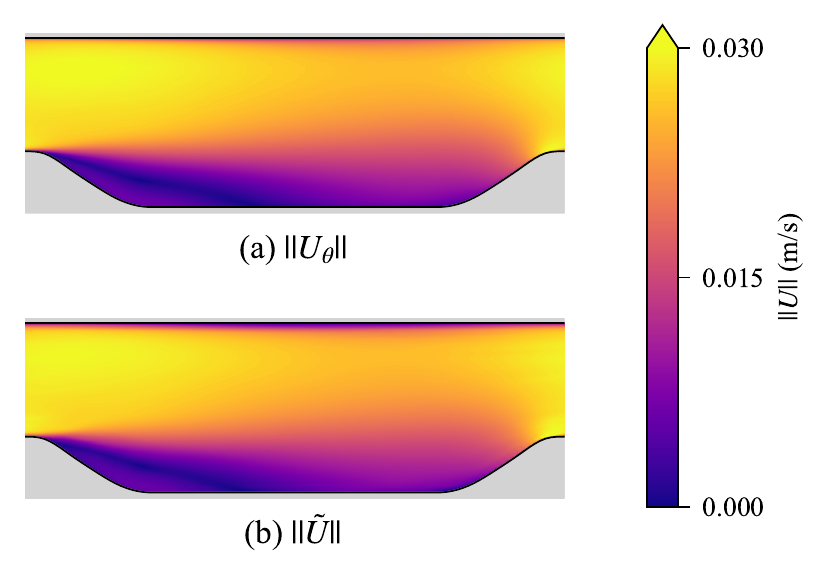}
\caption{\label{fig:U_mag_contours} \textit{A posteriori} contours of velocity magnitude after injecting the model predictions into the RANS equations: (a) DNS flow field and (b) result predicted by the present study. Here, $\| \tilde U \|$ is the magnitude of the velocity vector arising from the final converged solution, after injecting the model predictions for $\nu^\dagger_t$ and $a^\perp$.}
\end{figure}

Figure~\ref{fig:U_mag_contours} shows the contours of the velocity magnitude for the converged solution. The converged results agree well with the flow fields coming from the highly resolved DNS simulation. In particular, the size of the separated region is well predicted, as well as several other important features of the flow: namely, boundary layers along the top and bottom wall, reattachment of the flow, and acceleration over the rightward hill. 

\begin{figure*}
\includegraphics[]{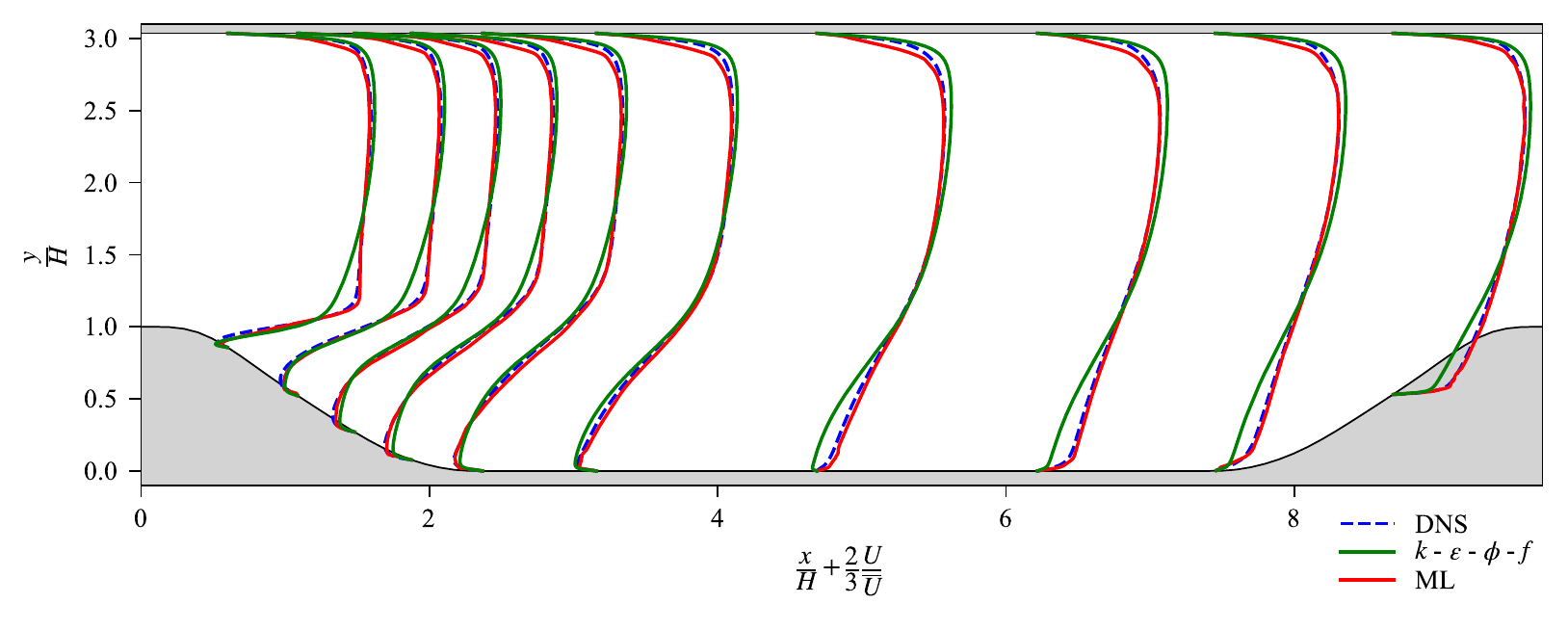}
\caption{\label{fig:U_samples} \textit{A posteriori} Samples of the $U$ velocity component along several lines throughout the flow field.}
\end{figure*}

\begin{figure*}
\includegraphics[]{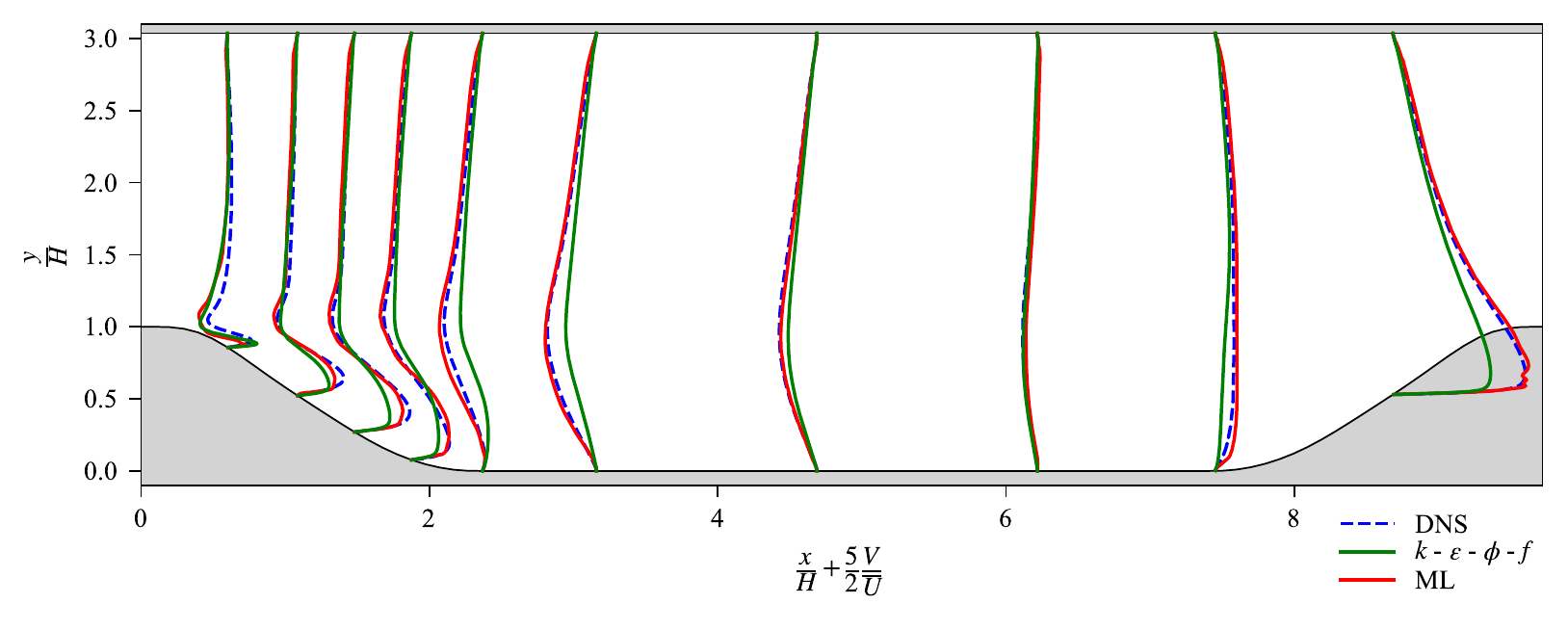}
\caption{\label{fig:V_samples} \textit{A posteriori} Samples of the $V$ velocity component along several lines throughout the flow field.}
\end{figure*}

Examining several slices of the velocity field in Figs~\ref{fig:U_samples} and \ref{fig:V_samples}, we see that both the $U$ and $V$ velocity components are captured well compared to the DNS fields. While the base RANS model (the green line in Figs~\ref{fig:U_samples} and \ref{fig:V_samples}) fails to predict the sharp transition in the $U$ profiles above the separated region, the ML augmented solution predicts these transitions accurately. Furthermore, the $U$ profiles after reattachment show much better agreement with the DNS solution than the base RANS model. Along the top wall, the ML augmented solution slightly underpredicts the $U$ profile, while the base RANS model overpredicts the $U$ profile.

While the test flow here is dominated by the streamwise flow, the $V$ profiles in Fig.~\ref{fig:V_samples} shed light on key areas that ML augmentation improves the solution. The base RANS models under-predicts the bulge in $V$ within the separated region, thereby indicating an over-prediction of separation. The tendency of commonly used RANS models to over-predict separation is widely known. Another key area of improvement in the solution concerns the acceleration of the flow over the right hill---the RANS model greatly under-predicts the $V$ component along the right hill, while the corrected solution exhibits a much more accurate estimate compared to DNS.


\begin{figure}
\includegraphics[]{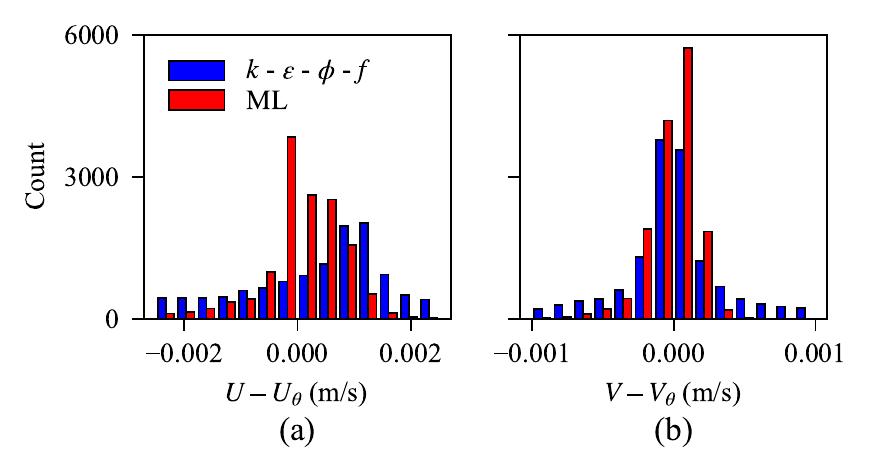}
\caption{\label{fig:U_V_error_histograms} Error distribution for the $U$ and $V$ velocity components after injecting the neural network model predictions for $\nu_t^\dagger$ and $a^\perp$ into the RANS equations: (a) error in the $U$ velocity component and (b) error in the $V$ velocity component.}
\end{figure}

Figure \ref{fig:U_V_error_histograms} compares the error distributions in $U$ and $V$ from the base RANS model to the ML augmented solution. The base RANS model displays a tendency to over-predict the $U$ component, demonstrated by a left skewed $U$ error histogram. While the augmented solution contains smaller error magnitudes, it also contains this same skewed tendency. The $V$ histograms are both roughly symmetric, with the augmented solution being slightly more skewed than the base RANS model.

\begin{figure}
\includegraphics[]{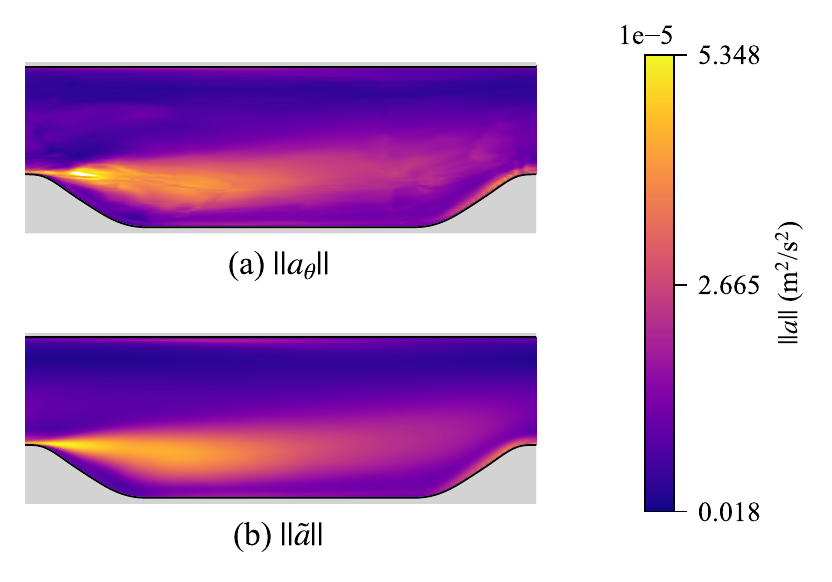}
\caption{\label{fig:a_mag_contours} \textit{A posteriori} contours of the anisotropy tensor magnitude: (a) ground-truth value obtained from DNS and (b) value predicted by the present study.}
\end{figure}

Since the linear part of $a$ continues to evolve as the solution converges, the full anisotropy tensor is an {\em a posteriori} quantity. Figure~\ref{fig:a_mag_contours} compares the predicted anisotropy tensor magnitude to the DNS value. The prediction for $a$ generally agrees well with $a_\theta$. A slight increase in the magnitude of $a$ as the flow accelerates over the right hill is captured, as well as the large increase in $a$ as the flow separates over the left hill. This accurate prediction of $a$ confirms the assumption in the present open loop framework that if $a^\perp$ and $\nu^\dagger_t$ are accurately predicted, $\vec{U}$ can evolve such that $\vec{U}\approx\vec{U}_\theta$, and therefore $S\approx S_\theta$. Finally, given $a^\perp \approx a^\perp_\theta$, $\nu^\dagger_t \approx \nu^\dagger_{t\theta}$, and $S\approx S_\theta$, we have $a \approx a_\theta$, which is the result shown in Fig.~\ref{fig:a_mag_contours}.

\begin{figure}
\includegraphics[]{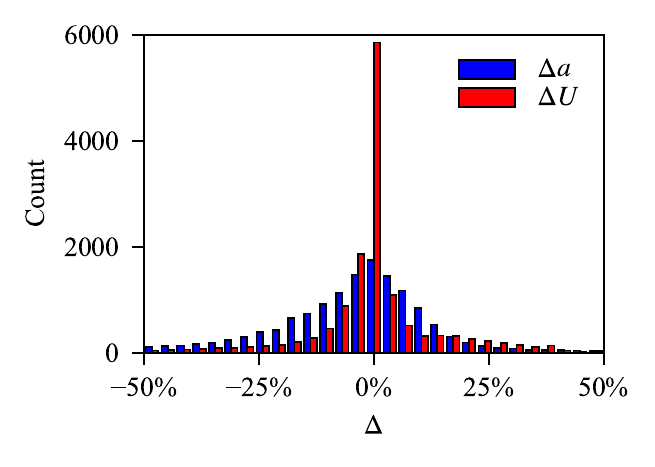}
\caption{\label{fig:Rel_error_comparison_hist} \textit{A posteriori} comparison of conditioning error in $U$ to conditioning errors in $a$. Relative error is calculated in an identical manner to Fig.~\ref{fig:Eddy_viscosity_error_a_theta_histograms}.}
\end{figure}

The conditioning problem for data-driven closures is often described as an amplification of errors in the closure term. It was of interest to determine whether this amplification was present in the solution after correcting $\nu^\dagger_t$ and $a^\perp$. Figure~\ref{fig:Rel_error_comparison_hist} compares the relative errors in the converged velocity field to the relative errors in the closure term, $a$. The errors in $U$ are \textit{reduced} compared to the errors in $a$. This reduction indicates that a well-conditioned closure formulation can also benefit the solution by suppressing errors in the closure term, leaving more room for errors in the machine learning model predictions. Clearly, the issue of conditioning deserves more attention in data-driven turbulence modelling. We recommend that a similar conditioning analysis be completed each time a new decomposition of the closure term is presented, or each time a new injection framework is proposed. 

\begin{figure}
\includegraphics[]{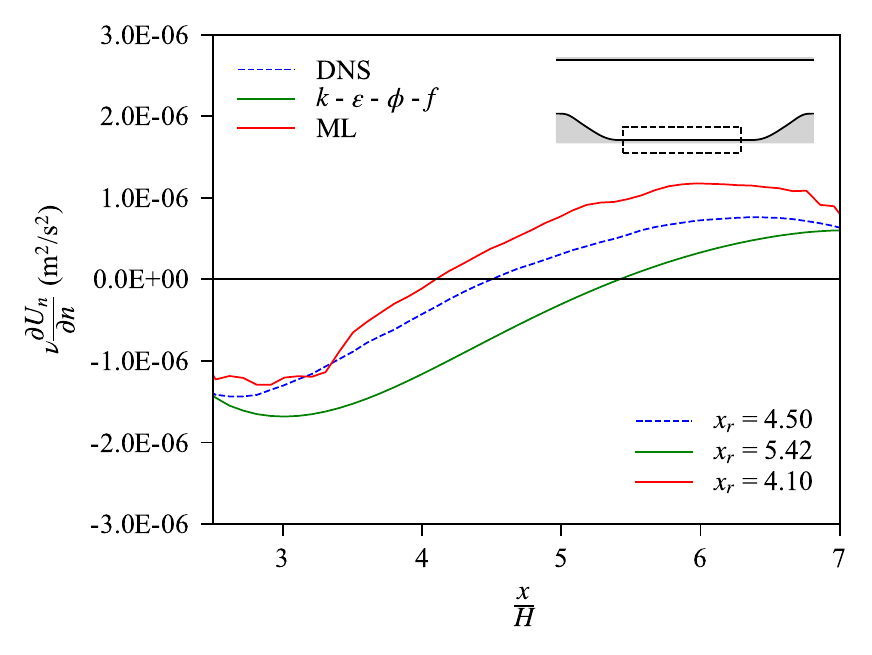}
\caption{\label{fig:WSS} Wall shear stress for a portion of the bottom wall. The area considered is shown in the top right corner of the figure. The reattachment position (i.e., the location at which the wall shear stress changes sign) is summarized in the bottom right corner.}
\end{figure}

An important flow phenomenon in the periodic hills test case is separation. For the $\alpha=1.2$ test case, the flow completely separates over the left hill, and reattaches along the bottom wall. The reattachment point is identified by the change in sign of the wall shear stress $\nu \frac{\partial U_n}{\partial n}$, where $n$ is the wall-normal direction, and $U_n$ is the velocity parallel to the wall. Figure~\ref{fig:WSS} shows the predicted wall shear stress along the reattaching region of the bottom wall.

The reattachment point $x_r$ in Fig.~\ref{fig:WSS} is the location where the wall shear stress changes signs. The base RANS model predicts a delayed reattachment point, a result of over-predicting the separation. The ML corrected velocity field reattaches slightly earlier than the DNS field. Nevertheless, the predicted reattachment point is much closer after applying the ML augmentation. The relative error in $x_r$ for the base RANS model is 20\%, while the error in the ML corrected solution is significantly reduced to 8.9\%.


\subsection{Interpreting the data-driven corrective closure model}\label{sec:interpret}
Given the rich input feature set used by the two neural networks in the present work, it was of interest to determine which features were found to be important by the model. In deep learning, this analysis is referred to as "interpretability"---taking a look inside the neural network black box. For this purpose, Shapley additive explanation (SHAP) values are commonly used to interpret machine learning models. The SHAP package by Lundberg and Lee~\cite{Lundberg2017} was used to collect the SHAP values for a range of $\nu^\dagger_t$ predictions by the EVNN. The EVNN was selected for an interpretability analysis, because it is the principle neural network in the present model. The eddy viscosity prediction is a critical injected component. SHAP values are used to determine the relative incremental contribution of each input feature on the prediction, while accounting for non-linear interactions between features. 

\begin{figure}
\includegraphics[]{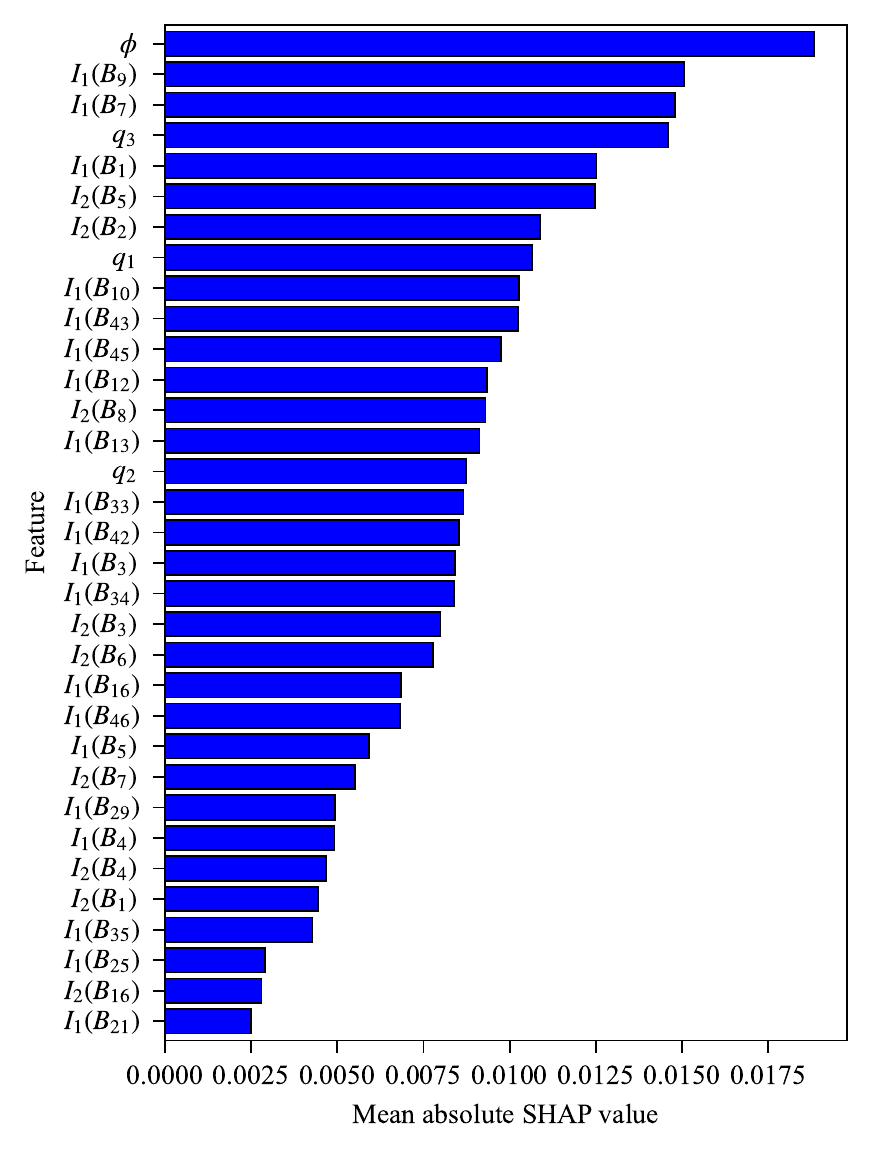}
\caption{\label{fig:shap_hist} Mean absolute SHAP values for each feature over the entire test case.  This value is used to measure the importance of each feature. The values are sorted from highest relevance ($\phi$) to lowest relevance ($I_1(B_{21})$). The SHAP values in this plot were calculated using the EVNN.}
\end{figure}

Summation of the SHAP values for all predictions can be used to determine the global relative importance of each input feature. Figure~\ref{fig:shap_hist} shows the relative importance of each input feature used in the EVNN. On a global basis, the use of $\phi$ as an input feature (made possible by the use of the $\phi$-$f$ model) was highly valuable, as the EVNN ranked it as the most important feature. Furthermore, the heuristic scalars ($q_1$, $q_2$, $q_3$) also ranked relatively high in terms of importance. The rest of the input features consists of various combinations of the strain rate, rotation rate, $k$ gradient, and $v^2$ gradient. The additional invariant $I_2$ used to extract additional features in the present work also added useful input features. Interestingly, there is no general correlation between the usefulness of $I_1$ and $I_2$ for a given basis tensor---the feature $I_1(B_7)$ ranked third, but $I_2(B_7)$ ranked twenty-fifth. The opposite is true for $B_5$---the second invariant ranked much higher than the first invariant.

\begin{figure}[!h]
\includegraphics[]{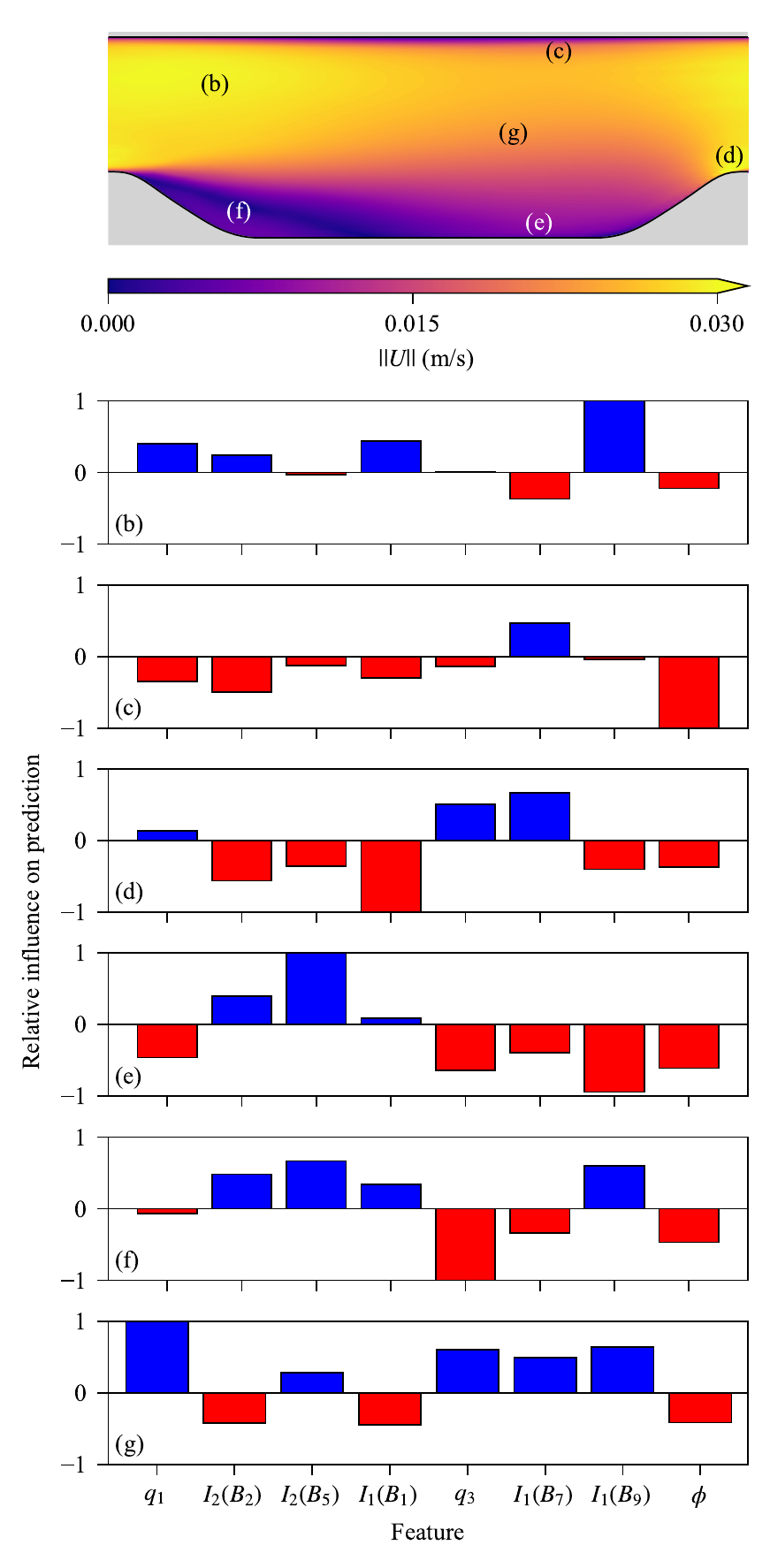}
\caption{\label{fig:local_shap} Relative feature importance at six locations in the flow: (a) locations in the flow domain and (b)--(g) bar plots showing the relative feature importance for the eight most important features in Fig.~\ref{fig:shap_hist}. The relative influence on the prediction is calculated by normalizing the SHAP values by the highest SHAP value within the top eight features at the given location.}
\end{figure}

Figure~\ref{fig:local_shap} shows the relative local SHAP values for the EVNN at several important points in the flow. The SHAP values for the eight most important features from Fig.~\ref{fig:shap_hist}, normalized by the maximum value at each location, are exhibited in this figure. The locations are annotated in Fig.~\ref{fig:local_shap}(a), and are summarized as follows: (b) the bulk flow region above the separated region; (c) the upper wall above the reattached and developing flow; (d) the accelerating flow above the right hill; (e) the lower wall just after reattachment; and, (f) the separated region. A blue colour indicates a positive influence on the $\nu^\dagger_t$ prediction, while a red colour indicates a negative influence. The sign indicates the direction each input feature "nudges" the output prediction, while the magnitude indicates the amount of influence.

Figure~\ref{fig:local_shap} shows that while some regions of the flow may feature similar phenomena (e.g., (b) and (g)), the relative importance of each feature is distinct. Though locations (c), (d), and (e) are all near the boundary layer, the magnitude of the $q_1$ importance is relatively low. Though $q_1$ is an input feature designed to detect near-wall regions, $q_1$ does not strongly influence $\tilde \nu^\dagger_t$ near the wall. At the top wall (c), the anisotropy measure $\phi$ is the most important feature. On the contrary, as the flow develops along the bottom wall at (e), $\phi$ is less important than some higher-order gradients of the mean field. In the separated region (f), the ratio of total to normal Reynolds stress ($q_3$) is significantly more important than the other input features. This observation justifies the heuristic basis for including $q_3$. As the flow recirculates along the bottom wall, it is possible that the EVNN has learned to associate a particular change in $q_3$ with separation. The anisotropy measure $\phi$ was not found to be as important in the separated region, but was important for the near-wall points (c) and (e).




\section{Conclusion}\label{sec:conclusion}
Applying machine learning to develop data-driven turbulence closures is a promising method to improve the accuracy of RANS simulations. However, a major issue for such models is the conditioning of the RANS equations after injecting a prediction for the closure term. If the framework is ill-conditioned, any errors made in the closure term could be amplified, and result in large errors in the converged mean field.

The decomposition of $\tau$ (and, in turn, $a$) proposed in this work was shown to have good conditioning, through an analysis similar to Brener {\em et al.}'s~\cite{Brener2021}. By injecting the training labels (calculated from DNS), rather than the model predictions, the resulting mean field provides the "upper limit" achievable at testing time. We demonstrated that decomposing $a$ into an optimal eddy viscosity ($\nu^\dagger_t$) and a remaining non-linear part ($a^\perp$) produced a well-conditioned closure. We also motivated the inclusion of the non-linear part of $a$ through evaluating the best prediction by a purely linear model.

We impose the often relaxed requirement of Galilean invariance on all input features. The input feature set was augmented by using an additional invariant to the one previously used in the literature. The minimal integrity basis for the strain rate, rotation rate, $k$ gradient, and $p$ gradient consists of 47 tensors, and the first invariant yields a scalar for each of these basis tensors. For a complex three-dimensional flow, using the second tensor invariant can effectively double the number of input features arising from this tensor basis to 94. For the two-dimensional flow considered here, we showed that many of these basis tensors are zero (Appendix~\ref{ap:invariants}), an outcome not previously discussed in detail. However, the input feature set used in the present work is still one of the richest feature sets applied for data-driven turbulence modelling to date. To determine which input features were found to be important, an interpretability framework (SHAP) was also applied to explain the predictions for the optimal eddy viscosity. Using this framework, it was found that the three heuristic scalars $q_1$, $q_2$ and $q_3$ ranked high along with $\phi$ (normalized wall-normal Reynolds stress) for the prediction of the optimal eddy viscosity.

Using one neural network to predict $\nu^\dagger_t$ and another to predict $a^\perp$, we demonstrated that these quantities are predictable with reasonable accuracy for a periodic hills dataset. Even after introducing model error into the predictions, the resulting mean fields agree well with the DNS results. {\em A posteriori} predictions of both the velocity field and the anisotropy tensor were demonstrated to agree well with the DNS results. Furthermore, the framework proposed here has a \textit{smaller} error in the velocity field than the closure prediction. This result is significant because it shows that under certain conditions, the RANS equations can suppress rather than amplify errors in the closure prediction.

In the present work, we also demonstrated that machine learning augmentation can be used to improve even a sophisticated RANS model. The base turbulence model used was the $\phi$-$f$ model, a type of $v^2$-$f$ model with three transport equations and one elliptic relaxation equation. The majority of previous studies has focused on simpler turbulence closure models, such as the standard $k$-$\varepsilon$ model or the $k$-$\omega$ model. This result indicates that machine learning augmentation has the potential to improve the results from even more sophisticated and recent RANS models, such as the elliptic blending (EB) $k$-$\varepsilon$-lag model~\cite{Lardeau2016}. Since the field $\phi$ was found to be a highly useful feature, we conclude that more sophisticated models may benefit even more from machine learning augmentation than the simpler RANS models. The wider set of mean fields for forming a complex closure relationship may provide a richer description of the flow, yielding more information for a machine learning model to use in predicting quantities of interest. This idea could be extended to develop an augmented Reynolds stress transport model, where an even broader set of mean fields are available.

Future work includes further optimizing the non-linear part of the anisotropy tensor $a^\perp$. While the present work imposes strict invariance quantities for the input features, the direct prediction of the components of $a^\perp$ means that the outputs do not possess the same invariance quantities. Possible options for remedying this discrepancy include using a tensor basis neural network (TBNN) for $a^\perp$ to construct neural network for $a^\perp$ that is invariant with respect to coordinate transformations (e.g., rotations, reflections). Further work also includes applying this framework to a broader set of flows. While the periodic hills dataset used here sufficiently demonstrated the merits of this framework, using a wider range of flows could provide additional useful information, especially when interpreting the closure model. For example, an analysis of the importance of various features for distinct flows could guide the input feature selection for a sophisticated three-dimensional machine learning closure model.

\begin{acknowledgments}
R.M. is supported by the Ontario Graduate Scholarship (OGS) program, and the Natural Sciences and Engineering Research Council of Canada (NSERC). The computational resources for this work were supported by the Tyler Lewis Clean Energy Research Foundation (TLCERF) and the Shared Hierarchical Academic Research Computing Network (SHARCNET).
\end{acknowledgments}

\section*{Data Availability Statement}



The data that support the findings of this study are openly available in "A curated dataset for data-driven turbulence modelling" at doi.org/10.34740/kaggle/dsv/2637500.
\appendix

\section{Invariants of the basis tensors}
\label{ap:invariants}
The input features used in the present work primarily consist of a set of invariants for a minimal integrity basis constructed from four mean field gradient tensors. Note that the flow considered in the present work is two-dimensional, which means that both the $z$-velocity component and all $z$-direction gradients are zero. As a result, the mean velocity gradient tensor is expressed as
\begin{equation}
\nabla \vec U = \begin{bmatrix}
\frac{\partial U}{\partial x} & \frac{\partial U}{\partial y} & 0\\
\frac{\partial V}{\partial x} & \frac{\partial V}{\partial y} & 0\\
0 & 0 & 0
\end{bmatrix}\ .
\end{equation}
The mean strain-rate tensor $S$ assumes the following form:
\begin{equation}
S = \frac{1}{2}\left(\nabla \vec U +\nabla {\vec U}^{\text{T}}\right)=\frac{1}{2}\begin{bmatrix}
2\frac{\partial U}{\partial x} & \frac{\partial U}{\partial y} + \frac{\partial V}{\partial x} & 0\\
\frac{\partial V}{\partial x}+\frac{\partial U}{\partial y} & 2\frac{\partial V}{\partial y} & 0\\
0 & 0 & 0
\end{bmatrix}\ .
\end{equation}
The mean rotation-rate tensor $R$ is given by
\begin{equation}
R = \frac{1}{2}\left(\nabla \vec U -\nabla {\vec U}^{\text{T}}\right)=\frac{1}{2}\begin{bmatrix}
0 & \frac{\partial U}{\partial y} - \frac{\partial V}{\partial x} & 0\\
\frac{\partial V}{\partial x}-\frac{\partial U}{\partial y} & 0 & 0\\
0 & 0 & 0
\end{bmatrix}\ .
\end{equation}
The TKE gradient vector can be recast as the following antisymmetric tensor:
\begin{equation}
A_k = \begin{bmatrix}
0 & 0 & \frac{\partial k}{\partial y}\\
0 & 0 & \frac{\partial k}{\partial x}\\
-\frac{\partial k}{\partial y} & -\frac{\partial k}{\partial x} & 0
\end{bmatrix}\ .
\end{equation}
Similarly, the gradient of $v^2$ (vector) can be re-expressed as the following antisymmetric tensor:
\begin{equation}
A_{v2} = \begin{bmatrix}
0 & 0 & \frac{\partial v^2}{\partial y}\\
0 & 0 & \frac{\partial v^2}{\partial x}\\
-\frac{\partial v^2}{\partial y} & -\frac{\partial v^2}{\partial x} & 0
\end{bmatrix}\ .
\end{equation}

\begin{table}[!h]
\caption{Basis tensors in the minimal integrity basis for $S$, $R$, $A_k$, and $A_{v2}$. For any tensor A, the first invariant is given by $I_1(A)=\text{tr}(A)$ and the second invariant is given by $I_2(A)=\tfrac{1}{2}[(\text{tr}(A))^2 - \text{tr}(A^2)]$.}
\label{tbl:invariants}
\begin{tabular}{cccccc}
\hline
\begin{tabular}[c]{@{}c@{}}Basis\\ tensor\end{tabular} & Expression           & \begin{tabular}[c]{@{}c@{}}Non-zero\\ $I_1$\end{tabular} & \begin{tabular}[c]{@{}c@{}}Non-zero\\ $I_2$\end{tabular} & \begin{tabular}[c]{@{}c@{}}Reason for\\ zero $I_1$\end{tabular} & \begin{tabular}[c]{@{}c@{}}Reason for\\ zero $I_2$\end{tabular} \\ \hline
$B_1$                                                  & $S^2$                & $\checkmark$                                             & $\checkmark$                                             & ---                                                             &  ---                                                           \\
$B_2$                                                  & $S^3$                & ---                                                      & $\checkmark$                                             & Continuity                                                      & ---                                                             \\
$B_3$                                                  & $R^2$                & $\checkmark$                                             & $\checkmark$                                             & ---                                                             & ---                                                             \\
$B_4$                                                  & $A_{v2}^2$           & $\checkmark$                                             & $\checkmark$                                             & ---                                                             & ---                                                             \\
$B_5$                                                  & $A_k^2$              & $\checkmark$                                             & $\checkmark$                                             & ---                                                             & ---                                                             \\
$B_6$                                                  & $R^2S$               &  ---                                                     & $\checkmark$                                             & Continuity                                                      & ---                                                             \\
$B_7$                                                  & $R^2S^2$             & $\checkmark$                                             & $\checkmark$                                             & ---                                                             & ---                                                             \\
$B_8$                                                  & $R^2SRS^2$           & ---                                                      & $\checkmark$                                             & Directly                                                        & ---                                                             \\
$B_9$                                                  & $A_{v2}^2 S$         & $\checkmark$                                             & ---                                                      & ---                                                             & Directly                                                        \\
$B_{10}$                                               & $A_{v2}^2 S^2$       & $\checkmark$                                             & ---                                                      & ---                                                             & Directly                                                        \\
$B_{11}$                                               & $A_{v2}^2SA_{v2}S^2$ &  ---                                                     & ---                                                      & Directly                                                        & Directly                                                        \\
$B_{12}$                                               & $A_{k}^2 S$          & $\checkmark$                                             & ---                                                      & ---                                                             & Directly                                                        \\
$B_{13}$                                               & $A_{k}^2 S^2$        & $\checkmark$                                             & ---                                                      & ---                                                             & Directly                                                        \\
$B_{14}$                                               & $A_k^2SA_k S^2$      & ---                                                      &  ---                                                     & Directly                                                        & Directly                                                        \\
$B_{15}$                                               & $A_{v2}A_k$          & ---                                                      &  ---                                                     & Directly                                                        & Directly                                                        \\
$B_{16}$                                               & $A_{v2}A_k$          & $\checkmark$                                             & $\checkmark$                                             & ---                                                             &  ---                                                            \\
$B_{17}$                                               & $RA_k$               & ---                                                      & ---                                                      & Directly                                                        & Directly                                                        \\
$B_{18}$                                               & $RA_{v2}S$           & ---                                                      & ---                                                      & Directly                                                        & Directly                                                        \\
$B_{19}$                                               & $RA_{v2}S^2$         & ---                                                      & ---                                                      & Directly                                                        & Directly                                                        \\
$B_{20}$                                               & $R^2A_{v2}S$         & ---                                                      & ---                                                      & Directly                                                        & Directly                                                        \\
$B_{21}$                                               & $A_{v2}RS$           & $\checkmark$                                             & ---                                                      & ---                                                             & Directly                                                        \\
$B_{22}$                                               & $R^2A_{v2}S^2$       & ---                                                      & ---                                                      & Directly                                                        & Directly                                                        \\
$B_{23}$                                               & $A_{v2}RS^2$         & ---                                                      & ---                                                      & Continuity                                                      & Directly                                                        \\
$B_{24}$                                               & $R^2SA_{v2}S^2$      & ---                                                      & ---                                                      & Directly                                                        & Directly                                                        \\
$B_{25}$                                               & $A_{v2}^2SRS^2$      & $\checkmark$                                             & ---                                                      & ---                                                             & Directly                                                        \\
$B_{26}$                                               & $RA_{v2}S$           &  ---                                                     & ---                                                      & Directly                                                        & Directly                                                        \\
$B_{27}$                                               & $RA_{v2}S^2$         & ---                                                      & ---                                                      & Directly                                                        & Directly                                                        \\
$B_{28}$                                               & $R^2A_kS$            & ---                                                      & ---                                                      & Directly                                                        & Directly                                                        \\
$B_{29}$                                               & $A_{k}^2RS$          & $\checkmark$                                             & ---                                                      & ---                                                             & Directly                                                         \\
$B_{30}$                                               & $R^2A_kS^2$          & ---                                                      & ---                                                      & Directly                                                        & Directly                                                        \\
$B_{31}$                                               & $A_k^2RS^2$          & ---                                                      & ---                                                      & Continuity                                                      & Directly                                                        \\
$B_{32}$                                               & $R^2SA_kS^2$         & ---                                                      & ---                                                      & Directly                                                        & Directly                                                        \\
$B_{33}$                                               & $A_{k}SRS^2$         & $\checkmark$                                             & ---                                                      & ---                                                             & Directly                                                        \\
$B_{34}$                                               & $A_{v2}A_kS$         & $\checkmark$                                             & ---                                                      & ---                                                             & Directly                                                        \\
$B_{35}$                                               & $A_{v2}A_k S^2$      & $\checkmark$                                             & ---                                                      & ---                                                             & Directly                                                        \\
$B_{36}$                                               & $A_{v2}^2A_kS$       & ---                                                      & ---                                                      & Directly                                                        & Directly                                                        \\
$B_{37}$                                               & $A_k^2A_{v2}S$       & ---                                                      & ---                                                      & Directly                                                        & Directly                                                        \\
$B_{38}$                                               & $A_{v2}A_kS^2$       & ---                                                      & ---                                                      & Directly                                                        & Directly                                                        \\
$B_{39}$                                               & $A_k^2 A_{v2} S^2$   & ---                                                      & ---                                                      & Directly                                                        & Directly                                                        \\
$B_{40}$                                               & $A_{v2}SA_kS^2$      & ---                                                      & ---                                                      & Directly                                                        & Directly                                                        \\
$B_{41}$                                               & $A_k^2SA_{v2}S^2$    & ---                                                      & ---                                                      & Directly                                                        & Directly                                                        \\
$B_{42}$                                               & $RA_{v2}A_k$         & $\checkmark$                                             & ---                                                      & ---                                                             & Directly                                                        \\
$B_{43}$                                               & $RA_{v2}A_kS$        & $\checkmark$                                             & ---                                                      & ---                                                             & Directly                                                        \\
$B_{44}$                                               & $RA_kA_{v2}S$        & $\checkmark$                                             & ---                                                      & ---                                                             & Directly                                                        \\
$B_{45}$                                               & $RA_{v2}A_kS^2$      & $\checkmark$                                             & ---                                                      & ---                                                             & Directly                                                        \\
$B_{46}$                                               & $RA_kA_{v2}S^2$      & $\checkmark$                                             & ---                                                      & ---                                                             & Directly                                                        \\
$B_{47}$                                               & $RA_{v2}SA_kS^2$     & ---                                                      & ---                                                      & Directly                                                        & Directly                                                        \\ \hline
\end{tabular}
\end{table}

A minimal integrity basis was constructed from the elements of the four independent tensors $S$, $R$, $A_k$, and $A_{v2}$ using the procedure developed by Spencer and Rivlin~\cite{Spencer1962}, and recently used by Wu {\em et al.}~\cite{Wu2018}. The resulting integrity basis has 47 tensors, summarized in Table~\ref{tbl:invariants}.

Maple, a symbolic math toolbox, was used to evaluate the first and second invariants for each of these basis tensors. Many of the invariants are identically zero, either directly (mathematically), or as a result of the incompressibility of the flow. The latter physical constraint is expressed through the continuity equation which, for a two-dimensional flow, reduces to
\begin{equation}
    \frac{\partial U}{\partial x} + \frac{\partial V}{\partial y}= 0\ .
\end{equation}
The third tensor invariant of any of the basis tensors comprising the minimal integrity basis formed from $S$, $R$, $A_k$ and $A_{v2}$ (viz.,  $\text{det}(A)$ for any basis tensor $A$) is necessarily (and, trivially) zero. The latter follows directly from the fact that the determinant of each of the base (independent) tensors $S$, $R$, $A_k$ and $A_{v2}$, used to construct the minimal integrity basis, vanishes identically.

Table~\ref{tbl:invariants} summarizes the results of the symbolic analysis, which was used to guide the input feature selection in the present work. A check mark in the table indicates that the associated invariant of the basis tensor ($B_1$--$B_{47}$) is non-zero and, as a consequence, can be used potentially as an non-trivial input feature for the prediction of $\nu_t^\dagger$ and $a^\perp$ in EVNN and apNN, respectively.

\newpage
\bibliography{references}

\begin{thebibliography}{47}%
\makeatletter
\providecommand \@ifxundefined [1]{%
 \@ifx{#1\undefined}
}%
\providecommand \@ifnum [1]{%
 \ifnum #1\expandafter \@firstoftwo
 \else \expandafter \@secondoftwo
 \fi
}%
\providecommand \@ifx [1]{%
 \ifx #1\expandafter \@firstoftwo
 \else \expandafter \@secondoftwo
 \fi
}%
\providecommand \natexlab [1]{#1}%
\providecommand \enquote  [1]{``#1''}%
\providecommand \bibnamefont  [1]{#1}%
\providecommand \bibfnamefont [1]{#1}%
\providecommand \citenamefont [1]{#1}%
\providecommand \href@noop [0]{\@secondoftwo}%
\providecommand \href [0]{\begingroup \@sanitize@url \@href}%
\providecommand \@href[1]{\@@startlink{#1}\@@href}%
\providecommand \@@href[1]{\endgroup#1\@@endlink}%
\providecommand \@sanitize@url [0]{\catcode `\\12\catcode `\$12\catcode
  `\&12\catcode `\#12\catcode `\^12\catcode `\_12\catcode `\%12\relax}%
\providecommand \@@startlink[1]{}%
\providecommand \@@endlink[0]{}%
\providecommand \url  [0]{\begingroup\@sanitize@url \@url }%
\providecommand \@url [1]{\endgroup\@href {#1}{\urlprefix }}%
\providecommand \urlprefix  [0]{URL }%
\providecommand \Eprint [0]{\href }%
\providecommand \doibase [0]{http://dx.doi.org/}%
\providecommand \selectlanguage [0]{\@gobble}%
\providecommand \bibinfo  [0]{\@secondoftwo}%
\providecommand \bibfield  [0]{\@secondoftwo}%
\providecommand \translation [1]{[#1]}%
\providecommand \BibitemOpen [0]{}%
\providecommand \bibitemStop [0]{}%
\providecommand \bibitemNoStop [0]{.\EOS\space}%
\providecommand \EOS [0]{\spacefactor3000\relax}%
\providecommand \BibitemShut  [1]{\csname bibitem#1\endcsname}%
\let\auto@bib@innerbib\@empty
\bibitem [{\citenamefont {Witherden}\ and\ \citenamefont
  {Jameson}(2017)}]{Witherden2017}%
  \BibitemOpen
  \bibfield  {author} {\bibinfo {author} {\bibfnamefont {F.~D.}\ \bibnamefont
  {Witherden}}\ and\ \bibinfo {author} {\bibfnamefont {A.}~\bibnamefont
  {Jameson}},\ }\bibfield  {title} {\enquote {\bibinfo {title} {{Future
  directions of computational fluid dynamics}},}\ }\href@noop {} {\bibfield
  {journal} {\bibinfo  {journal} {23rd AIAA Computational Fluid Dynamics
  Conference, 2017}\ ,\ \bibinfo {pages} {1--16}} (\bibinfo {year}
  {2017})}\BibitemShut {NoStop}%
\bibitem [{\citenamefont {Slotnick}\ \emph {et~al.}(2014)\citenamefont
  {Slotnick}, \citenamefont {Khodadoust}, \citenamefont {Alonso},\ and\
  \citenamefont {Darmofal}}]{CFD2030}%
  \BibitemOpen
  \bibfield  {author} {\bibinfo {author} {\bibfnamefont {J.}~\bibnamefont
  {Slotnick}}, \bibinfo {author} {\bibfnamefont {A.}~\bibnamefont
  {Khodadoust}}, \bibinfo {author} {\bibfnamefont {J.}~\bibnamefont {Alonso}},
  \ and\ \bibinfo {author} {\bibfnamefont {D.}~\bibnamefont {Darmofal}},\
  }\href@noop {} {\enquote {\bibinfo {title} {{CFD Vision 2030 Study: A Path to
  Revolutionary Computational Aerosciences}},}\ }\bibinfo {type} {Tech. Rep.}\
  \bibinfo {number} {March}\ (\bibinfo {year} {2014})\BibitemShut {NoStop}%
\bibitem [{\citenamefont {Duraisamy}(2020)}]{Duraisamy2020}%
  \BibitemOpen
  \bibfield  {author} {\bibinfo {author} {\bibfnamefont {K.}~\bibnamefont
  {Duraisamy}},\ }\bibfield  {title} {\enquote {\bibinfo {title} {{Perspectives
  on machine learning-augmented Reynolds-averaged and large eddy simulation
  models of turbulence}},}\ }\href {http://arxiv.org/abs/2009.10675} {\bibfield
   {journal} {\bibinfo  {journal} {arXiv}\ ,\ \bibinfo {pages} {1--25}}
  (\bibinfo {year} {2020})}\BibitemShut {NoStop}%
\bibitem [{\citenamefont {Duraisamy}, \citenamefont {Iaccarino},\ and\
  \citenamefont {Xiao}(2019)}]{Duraisamy2019a}%
  \BibitemOpen
  \bibfield  {author} {\bibinfo {author} {\bibfnamefont {K.}~\bibnamefont
  {Duraisamy}}, \bibinfo {author} {\bibfnamefont {G.}~\bibnamefont
  {Iaccarino}}, \ and\ \bibinfo {author} {\bibfnamefont {H.}~\bibnamefont
  {Xiao}},\ }\bibfield  {title} {\enquote {\bibinfo {title} {{Turbulence
  Modeling in the Age of Data}},}\ }\href {\doibase
  10.1146/annurev-fluid-010518-040547} {\bibfield  {journal} {\bibinfo
  {journal} {Annual Review of Fluid Mechanics}\ }\textbf {\bibinfo {volume}
  {51}},\ \bibinfo {pages} {357--377} (\bibinfo {year} {2019})}\BibitemShut
  {NoStop}%
\bibitem [{\citenamefont {Brunton}, \citenamefont {Noack},\ and\ \citenamefont
  {Koumoutsakos}(2020)}]{Brunton2020}%
  \BibitemOpen
  \bibfield  {author} {\bibinfo {author} {\bibfnamefont {S.~L.}\ \bibnamefont
  {Brunton}}, \bibinfo {author} {\bibfnamefont {B.~R.}\ \bibnamefont {Noack}},
  \ and\ \bibinfo {author} {\bibfnamefont {P.}~\bibnamefont {Koumoutsakos}},\
  }\bibfield  {title} {\enquote {\bibinfo {title} {{Machine Learning for Fluid
  Mechanics}},}\ }\href {\doibase 10.1146/annurev-fluid-010719-060214}
  {\bibfield  {journal} {\bibinfo  {journal} {Annual Review of Fluid
  Mechanics}\ }\textbf {\bibinfo {volume} {52}},\ \bibinfo {pages} {1--32}
  (\bibinfo {year} {2020})}\BibitemShut {NoStop}%
\bibitem [{\citenamefont {Kutz}(2017)}]{Kutz2017}%
  \BibitemOpen
  \bibfield  {author} {\bibinfo {author} {\bibfnamefont {J.~N.}\ \bibnamefont
  {Kutz}},\ }\bibfield  {title} {\enquote {\bibinfo {title} {{Deep learning in
  fluid dynamics}},}\ }\href {\doibase 10.1017/jfm.2016.803} {\bibfield
  {journal} {\bibinfo  {journal} {Journal of Fluid Mechanics}\ }\textbf
  {\bibinfo {volume} {814}},\ \bibinfo {pages} {1--4} (\bibinfo {year}
  {2017})}\BibitemShut {NoStop}%
\bibitem [{\citenamefont {Wang}, \citenamefont {Wu},\ and\ \citenamefont
  {Xiao}(2017)}]{Wang2017}%
  \BibitemOpen
  \bibfield  {author} {\bibinfo {author} {\bibfnamefont {J.~X.}\ \bibnamefont
  {Wang}}, \bibinfo {author} {\bibfnamefont {J.~L.}\ \bibnamefont {Wu}}, \ and\
  \bibinfo {author} {\bibfnamefont {H.}~\bibnamefont {Xiao}},\ }\bibfield
  {title} {\enquote {\bibinfo {title} {{Physics-informed machine learning
  approach for reconstructing Reynolds stress modeling discrepancies based on
  DNS data}},}\ }\href {\doibase 10.1103/PhysRevFluids.2.034603} {\bibfield
  {journal} {\bibinfo  {journal} {Physical Review Fluids}\ }\textbf {\bibinfo
  {volume} {2}},\ \bibinfo {pages} {1--22} (\bibinfo {year}
  {2017})}\BibitemShut {NoStop}%
\bibitem [{\citenamefont {Duraisamy}(2021)}]{Duraisamy2021}%
  \BibitemOpen
  \bibfield  {author} {\bibinfo {author} {\bibfnamefont {K.}~\bibnamefont
  {Duraisamy}},\ }\bibfield  {title} {\enquote {\bibinfo {title} {{Perspectives
  on machine learning-augmented Reynolds-averaged and large eddy simulation
  models of turbulence}},}\ }\href {\doibase 10.1103/PhysRevFluids.6.050504}
  {\bibfield  {journal} {\bibinfo  {journal} {Physical Review Fluids}\ }\textbf
  {\bibinfo {volume} {6}},\ \bibinfo {pages} {1--16} (\bibinfo {year}
  {2021})}\BibitemShut {NoStop}%
\bibitem [{\citenamefont {Zhu}\ and\ \citenamefont {Dinh}(2020)}]{Zhu2020}%
  \BibitemOpen
  \bibfield  {author} {\bibinfo {author} {\bibfnamefont {Y.}~\bibnamefont
  {Zhu}}\ and\ \bibinfo {author} {\bibfnamefont {N.}~\bibnamefont {Dinh}},\
  }\bibfield  {title} {\enquote {\bibinfo {title} {{A data-driven approach for
  turbulence modeling}},}\ }\href {http://arxiv.org/abs/2005.00426} {\bibfield
  {journal} {\bibinfo  {journal} {arXiv}\ } (\bibinfo {year}
  {2020})}\BibitemShut {NoStop}%
\bibitem [{\citenamefont {Chang}\ and\ \citenamefont {Dinh}(2018)}]{Chang2018}%
  \BibitemOpen
  \bibfield  {author} {\bibinfo {author} {\bibfnamefont {C.-W.}\ \bibnamefont
  {Chang}}\ and\ \bibinfo {author} {\bibfnamefont {N.~T.}\ \bibnamefont
  {Dinh}},\ }\bibfield  {title} {\enquote {\bibinfo {title} {{Reynolds-Averaged
  Turbulence Modeling Using Type I and Type II Machine Learning Frameworks with
  Deep Learning}},}\ }\href {http://arxiv.org/abs/1804.01065} {\  (\bibinfo
  {year} {2018})}\BibitemShut {NoStop}%
\bibitem [{\citenamefont {Jiang}(2021)}]{Jiang2021}%
  \BibitemOpen
  \bibfield  {author} {\bibinfo {author} {\bibfnamefont {C.}~\bibnamefont
  {Jiang}},\ }\bibfield  {title} {\enquote {\bibinfo {title} {{Physics-guided
  deep learning framework for predictive modeling of Reynolds stress
  anisotropy}},}\ }\href {http://arxiv.org/abs/2102.03767} {\ ,\ \bibinfo
  {pages} {1--46} (\bibinfo {year} {2021})}\BibitemShut {NoStop}%
\bibitem [{\citenamefont {Bhushan}\ \emph {et~al.}(2021)\citenamefont
  {Bhushan}, \citenamefont {Burgreen}, \citenamefont {Brewer},\ and\
  \citenamefont {Dettwiller}}]{Bhushan2021}%
  \BibitemOpen
  \bibfield  {author} {\bibinfo {author} {\bibfnamefont {S.}~\bibnamefont
  {Bhushan}}, \bibinfo {author} {\bibfnamefont {G.~W.}\ \bibnamefont
  {Burgreen}}, \bibinfo {author} {\bibfnamefont {W.}~\bibnamefont {Brewer}}, \
  and\ \bibinfo {author} {\bibfnamefont {I.~D.}\ \bibnamefont {Dettwiller}},\
  }\bibfield  {title} {\enquote {\bibinfo {title} {{Development and validation
  of a machine learned turbulence model}},}\ }\href {\doibase
  10.3390/en14051465} {\bibfield  {journal} {\bibinfo  {journal} {Energies}\
  }\textbf {\bibinfo {volume} {14}} (\bibinfo {year} {2021}),\
  10.3390/en14051465}\BibitemShut {NoStop}%
\bibitem [{\citenamefont {Liu}\ \emph {et~al.}(2021)\citenamefont {Liu},
  \citenamefont {Fang}, \citenamefont {Rolfo}, \citenamefont {Moulinec},\ and\
  \citenamefont {Emerson}}]{Liu2021}%
  \BibitemOpen
  \bibfield  {author} {\bibinfo {author} {\bibfnamefont {W.}~\bibnamefont
  {Liu}}, \bibinfo {author} {\bibfnamefont {J.}~\bibnamefont {Fang}}, \bibinfo
  {author} {\bibfnamefont {S.}~\bibnamefont {Rolfo}}, \bibinfo {author}
  {\bibfnamefont {C.}~\bibnamefont {Moulinec}}, \ and\ \bibinfo {author}
  {\bibfnamefont {D.~R.}\ \bibnamefont {Emerson}},\ }\bibfield  {title}
  {\enquote {\bibinfo {title} {{An iterative machine-learning framework for
  RANS turbulence modeling}},}\ }\href {\doibase
  10.1016/j.ijheatfluidflow.2021.108822} {\bibfield  {journal} {\bibinfo
  {journal} {International Journal of Heat and Fluid Flow}\ }\textbf {\bibinfo
  {volume} {90}},\ \bibinfo {pages} {108822} (\bibinfo {year}
  {2021})}\BibitemShut {NoStop}%
\bibitem [{\citenamefont {Srinivasan}\ \emph {et~al.}(2019)\citenamefont
  {Srinivasan}, \citenamefont {Guastoni}, \citenamefont {Azizpour},
  \citenamefont {Schlatter},\ and\ \citenamefont {Vinuesa}}]{Srinivasan2019}%
  \BibitemOpen
  \bibfield  {author} {\bibinfo {author} {\bibfnamefont {P.~A.}\ \bibnamefont
  {Srinivasan}}, \bibinfo {author} {\bibfnamefont {L.}~\bibnamefont
  {Guastoni}}, \bibinfo {author} {\bibfnamefont {H.}~\bibnamefont {Azizpour}},
  \bibinfo {author} {\bibfnamefont {P.}~\bibnamefont {Schlatter}}, \ and\
  \bibinfo {author} {\bibfnamefont {R.}~\bibnamefont {Vinuesa}},\ }\bibfield
  {title} {\enquote {\bibinfo {title} {{Predictions of turbulent shear flows
  using deep neural networks}},}\ }\href {\doibase
  10.1103/PhysRevFluids.4.054603} {\bibfield  {journal} {\bibinfo  {journal}
  {Physical Review Fluids}\ }\textbf {\bibinfo {volume} {4}},\ \bibinfo {pages}
  {1--15} (\bibinfo {year} {2019})}\BibitemShut {NoStop}%
\bibitem [{\citenamefont {Yin}\ \emph {et~al.}(2020)\citenamefont {Yin},
  \citenamefont {Yang}, \citenamefont {Zhang}, \citenamefont {Chen},\ and\
  \citenamefont {Fu}}]{Yin2020}%
  \BibitemOpen
  \bibfield  {author} {\bibinfo {author} {\bibfnamefont {Y.}~\bibnamefont
  {Yin}}, \bibinfo {author} {\bibfnamefont {P.}~\bibnamefont {Yang}}, \bibinfo
  {author} {\bibfnamefont {Y.}~\bibnamefont {Zhang}}, \bibinfo {author}
  {\bibfnamefont {H.}~\bibnamefont {Chen}}, \ and\ \bibinfo {author}
  {\bibfnamefont {S.}~\bibnamefont {Fu}},\ }\bibfield  {title} {\enquote
  {\bibinfo {title} {{Feature selection and processing of turbulence modeling
  based on an artificial neural network}},}\ }\href {\doibase
  10.1063/5.0022561} {\bibfield  {journal} {\bibinfo  {journal} {Physics of
  Fluids}\ }\textbf {\bibinfo {volume} {32}},\ \bibinfo {pages} {105117}
  (\bibinfo {year} {2020})}\BibitemShut {NoStop}%
\bibitem [{\citenamefont {Fang}\ \emph {et~al.}(2019)\citenamefont {Fang},
  \citenamefont {Sondak}, \citenamefont {Protopapas},\ and\ \citenamefont
  {Succi}}]{Fang2020}%
  \BibitemOpen
  \bibfield  {author} {\bibinfo {author} {\bibfnamefont {R.}~\bibnamefont
  {Fang}}, \bibinfo {author} {\bibfnamefont {D.}~\bibnamefont {Sondak}},
  \bibinfo {author} {\bibfnamefont {P.}~\bibnamefont {Protopapas}}, \ and\
  \bibinfo {author} {\bibfnamefont {S.}~\bibnamefont {Succi}},\ }\bibfield
  {title} {\enquote {\bibinfo {title} {{Neural Network Models for the
  Anisotropic Reynolds Stress Tensor in Turbulent Channel Flow}},}\ }\href
  {\doibase 10.1080/14685248.2019.1706742} {\bibfield  {journal} {\bibinfo
  {journal} {Journal of Turbulence}\ }\textbf {\bibinfo {volume} {21}},\
  \bibinfo {pages} {525--543} (\bibinfo {year} {2019})}\BibitemShut {NoStop}%
\bibitem [{\citenamefont {Zhang}\ \emph
  {et~al.}(2019{\natexlab{a}})\citenamefont {Zhang}, \citenamefont {Wu},
  \citenamefont {Coutier-Delgosha},\ and\ \citenamefont {Xiao}}]{Zhang2019b}%
  \BibitemOpen
  \bibfield  {author} {\bibinfo {author} {\bibfnamefont {X.}~\bibnamefont
  {Zhang}}, \bibinfo {author} {\bibfnamefont {J.}~\bibnamefont {Wu}}, \bibinfo
  {author} {\bibfnamefont {O.}~\bibnamefont {Coutier-Delgosha}}, \ and\
  \bibinfo {author} {\bibfnamefont {H.}~\bibnamefont {Xiao}},\ }\bibfield
  {title} {\enquote {\bibinfo {title} {{Recent progress in augmenting
  turbulence models with physics-informed machine learning}},}\ }\href
  {\doibase 10.1007/s42241-019-0089-y} {\bibfield  {journal} {\bibinfo
  {journal} {Journal of Hydrodynamics}\ }\textbf {\bibinfo {volume} {31}},\
  \bibinfo {pages} {1153--1158} (\bibinfo {year}
  {2019}{\natexlab{a}})}\BibitemShut {NoStop}%
\bibitem [{\citenamefont {Zhu}\ \emph {et~al.}(2019)\citenamefont {Zhu},
  \citenamefont {Zhang}, \citenamefont {Kou},\ and\ \citenamefont
  {Liu}}]{Zhu2019}%
  \BibitemOpen
  \bibfield  {author} {\bibinfo {author} {\bibfnamefont {L.}~\bibnamefont
  {Zhu}}, \bibinfo {author} {\bibfnamefont {W.}~\bibnamefont {Zhang}}, \bibinfo
  {author} {\bibfnamefont {J.}~\bibnamefont {Kou}}, \ and\ \bibinfo {author}
  {\bibfnamefont {Y.}~\bibnamefont {Liu}},\ }\bibfield  {title} {\enquote
  {\bibinfo {title} {{Machine learning methods for turbulence modeling in
  subsonic flows around airfoils}},}\ }\href {\doibase 10.1063/1.5061693}
  {\bibfield  {journal} {\bibinfo  {journal} {Physics of Fluids}\ }\textbf
  {\bibinfo {volume} {31}} (\bibinfo {year} {2019}),\
  10.1063/1.5061693}\BibitemShut {NoStop}%
\bibitem [{\citenamefont {Matai}\ and\ \citenamefont
  {Durbin}(2019)}]{Matai2019}%
  \BibitemOpen
  \bibfield  {author} {\bibinfo {author} {\bibfnamefont {R.}~\bibnamefont
  {Matai}}\ and\ \bibinfo {author} {\bibfnamefont {P.~A.}\ \bibnamefont
  {Durbin}},\ }\bibfield  {title} {\enquote {\bibinfo {title} {{Zonal Eddy
  Viscosity Models Based on Machine Learning}},}\ }\href {\doibase
  10.1007/s10494-019-00011-5} {\bibfield  {journal} {\bibinfo  {journal} {Flow,
  Turbulence and Combustion}\ }\textbf {\bibinfo {volume} {103}},\ \bibinfo
  {pages} {93--109} (\bibinfo {year} {2019})}\BibitemShut {NoStop}%
\bibitem [{\citenamefont {Nikolaou}\ \emph {et~al.}(2020)\citenamefont
  {Nikolaou}, \citenamefont {Chrysostomou}, \citenamefont {Minamoto},\ and\
  \citenamefont {Vervisch}}]{Nikolaou2020}%
  \BibitemOpen
  \bibfield  {author} {\bibinfo {author} {\bibfnamefont {Z.~M.}\ \bibnamefont
  {Nikolaou}}, \bibinfo {author} {\bibfnamefont {C.}~\bibnamefont
  {Chrysostomou}}, \bibinfo {author} {\bibfnamefont {Y.}~\bibnamefont
  {Minamoto}}, \ and\ \bibinfo {author} {\bibfnamefont {L.}~\bibnamefont
  {Vervisch}},\ }\bibfield  {title} {\enquote {\bibinfo {title} {{Evaluation of
  a Neural Network-Based Closure for the Unresolved Stresses in Turbulent
  Premixed V-Flames}},}\ }\href {\doibase 10.1007/s10494-020-00170-w}
  {\bibfield  {journal} {\bibinfo  {journal} {Flow, Turbulence and Combustion}\
  } (\bibinfo {year} {2020}),\ 10.1007/s10494-020-00170-w}\BibitemShut
  {NoStop}%
\bibitem [{\citenamefont {Tan}\ \emph {et~al.}(2021)\citenamefont {Tan},
  \citenamefont {He}, \citenamefont {Rigas},\ and\ \citenamefont
  {Vahdati}}]{Tan2021}%
  \BibitemOpen
  \bibfield  {author} {\bibinfo {author} {\bibfnamefont {J.}~\bibnamefont
  {Tan}}, \bibinfo {author} {\bibfnamefont {X.}~\bibnamefont {He}}, \bibinfo
  {author} {\bibfnamefont {G.}~\bibnamefont {Rigas}}, \ and\ \bibinfo {author}
  {\bibfnamefont {M.}~\bibnamefont {Vahdati}},\ }\bibfield  {title} {\enquote
  {\bibinfo {title} {{Towards Explainable Machine-Learning-Assisted Turbulence
  Modeling for Transonic Flows}},}\ }\href@noop {} {\ ,\ \bibinfo {pages}
  {1--12} (\bibinfo {year} {2021})}\BibitemShut {NoStop}%
\bibitem [{\citenamefont {Kaandorp}(2018)}]{Kaandorp2018}%
  \BibitemOpen
  \bibfield  {author} {\bibinfo {author} {\bibfnamefont {M.}~\bibnamefont
  {Kaandorp}},\ }\bibfield  {title} {\enquote {\bibinfo {title} {{Machine
  Learning for Data-Driven RANS Turbulence Modelling}},}\ }\href
  {https://repository.tudelft.nl/islandora/object/uuid%3Af833e151-7c0f-414c-8217-5af783c88474?collection=education}
  {\bibfield  {journal} {\bibinfo  {journal} {Delft University of Technology}\
  } (\bibinfo {year} {2018})}\BibitemShut {NoStop}%
\bibitem [{\citenamefont {Kaandorp}\ and\ \citenamefont
  {Dwight}(2020)}]{Kaandorp2020}%
  \BibitemOpen
  \bibfield  {author} {\bibinfo {author} {\bibfnamefont {M.~L.}\ \bibnamefont
  {Kaandorp}}\ and\ \bibinfo {author} {\bibfnamefont {R.~P.}\ \bibnamefont
  {Dwight}},\ }\bibfield  {title} {\enquote {\bibinfo {title} {{Data-driven
  modelling of the Reynolds stress tensor using random forests with
  invariance}},}\ }\href {\doibase 10.1016/j.compfluid.2020.104497} {\bibfield
  {journal} {\bibinfo  {journal} {Computers and Fluids}\ }\textbf {\bibinfo
  {volume} {202}},\ \bibinfo {pages} {104497} (\bibinfo {year}
  {2020})}\BibitemShut {NoStop}%
\bibitem [{\citenamefont {Song}\ \emph {et~al.}(2019)\citenamefont {Song},
  \citenamefont {Zhang}, \citenamefont {Wang}, \citenamefont {Ye},\ and\
  \citenamefont {Huang}}]{Song2019}%
  \BibitemOpen
  \bibfield  {author} {\bibinfo {author} {\bibfnamefont {X.~D.}\ \bibnamefont
  {Song}}, \bibinfo {author} {\bibfnamefont {Z.}~\bibnamefont {Zhang}},
  \bibinfo {author} {\bibfnamefont {Y.~W.}\ \bibnamefont {Wang}}, \bibinfo
  {author} {\bibfnamefont {S.~R.}\ \bibnamefont {Ye}}, \ and\ \bibinfo {author}
  {\bibfnamefont {C.~G.}\ \bibnamefont {Huang}},\ }\bibfield  {title} {\enquote
  {\bibinfo {title} {{Reconstruction of RANS model and cross-validation of flow
  field based on tensor basis neural network}},}\ }\href@noop {} {\bibfield
  {journal} {\bibinfo  {journal} {Proceedings of the ASME-JSME-KSME 2019 8th
  Joint Fluids Engineering Conference}\ ,\ \bibinfo {pages} {1--6}} (\bibinfo
  {year} {2019})}\BibitemShut {NoStop}%
\bibitem [{\citenamefont {Zhang}\ \emph
  {et~al.}(2019{\natexlab{b}})\citenamefont {Zhang}, \citenamefont {Song},
  \citenamefont {Ye}, \citenamefont {Wang}, \citenamefont {Huang},
  \citenamefont {An},\ and\ \citenamefont {Chen}}]{Zhang2019a}%
  \BibitemOpen
  \bibfield  {author} {\bibinfo {author} {\bibfnamefont {Z.}~\bibnamefont
  {Zhang}}, \bibinfo {author} {\bibfnamefont {X.~D.}\ \bibnamefont {Song}},
  \bibinfo {author} {\bibfnamefont {S.~R.}\ \bibnamefont {Ye}}, \bibinfo
  {author} {\bibfnamefont {Y.~W.}\ \bibnamefont {Wang}}, \bibinfo {author}
  {\bibfnamefont {C.~G.}\ \bibnamefont {Huang}}, \bibinfo {author}
  {\bibfnamefont {Y.~R.}\ \bibnamefont {An}}, \ and\ \bibinfo {author}
  {\bibfnamefont {Y.~S.}\ \bibnamefont {Chen}},\ }\bibfield  {title} {\enquote
  {\bibinfo {title} {{Application of deep learning method to Reynolds stress
  models of channel flow based on reduced-order modeling of DNS data}},}\
  }\href {\doibase 10.1007/s42241-018-0156-9} {\bibfield  {journal} {\bibinfo
  {journal} {Journal of Hydrodynamics}\ }\textbf {\bibinfo {volume} {31}},\
  \bibinfo {pages} {58--65} (\bibinfo {year} {2019}{\natexlab{b}})}\BibitemShut
  {NoStop}%
\bibitem [{\citenamefont {Ling}, \citenamefont {Kurzawski},\ and\ \citenamefont
  {Templeton}(2016)}]{Ling2016}%
  \BibitemOpen
  \bibfield  {author} {\bibinfo {author} {\bibfnamefont {J.}~\bibnamefont
  {Ling}}, \bibinfo {author} {\bibfnamefont {A.}~\bibnamefont {Kurzawski}}, \
  and\ \bibinfo {author} {\bibfnamefont {J.}~\bibnamefont {Templeton}},\
  }\bibfield  {title} {\enquote {\bibinfo {title} {{Reynolds averaged
  turbulence modelling using deep neural networks with embedded invariance}},}\
  }\href {\doibase 10.1017/jfm.2016.615} {\bibfield  {journal} {\bibinfo
  {journal} {Journal of Fluid Mechanics}\ }\textbf {\bibinfo {volume} {807}},\
  \bibinfo {pages} {155--166} (\bibinfo {year} {2016})}\BibitemShut {NoStop}%
\bibitem [{\citenamefont {Wu}, \citenamefont {Xiao},\ and\ \citenamefont
  {Paterson}(2018)}]{Wu2018}%
  \BibitemOpen
  \bibfield  {author} {\bibinfo {author} {\bibfnamefont {J.~L.}\ \bibnamefont
  {Wu}}, \bibinfo {author} {\bibfnamefont {H.}~\bibnamefont {Xiao}}, \ and\
  \bibinfo {author} {\bibfnamefont {E.}~\bibnamefont {Paterson}},\ }\bibfield
  {title} {\enquote {\bibinfo {title} {{Physics-informed machine learning
  approach for augmenting turbulence models: A comprehensive framework}},}\
  }\href {\doibase 10.1103/PhysRevFluids.3.074602} {\bibfield  {journal}
  {\bibinfo  {journal} {Physical Review Fluids}\ }\textbf {\bibinfo {volume}
  {7}},\ \bibinfo {pages} {1--28} (\bibinfo {year} {2018})}\BibitemShut
  {NoStop}%
\bibitem [{\citenamefont {Cruz}\ \emph {et~al.}(2019)\citenamefont {Cruz},
  \citenamefont {Thompson}, \citenamefont {Sampaio},\ and\ \citenamefont
  {Bacchi}}]{Cruz2019}%
  \BibitemOpen
  \bibfield  {author} {\bibinfo {author} {\bibfnamefont {M.~A.}\ \bibnamefont
  {Cruz}}, \bibinfo {author} {\bibfnamefont {R.~L.}\ \bibnamefont {Thompson}},
  \bibinfo {author} {\bibfnamefont {L.~E.}\ \bibnamefont {Sampaio}}, \ and\
  \bibinfo {author} {\bibfnamefont {R.~D.}\ \bibnamefont {Bacchi}},\ }\bibfield
   {title} {\enquote {\bibinfo {title} {{The use of the Reynolds force vector
  in a physics informed machine learning approach for predictive turbulence
  modeling}},}\ }\href {\doibase 10.1016/j.compfluid.2019.104258} {\bibfield
  {journal} {\bibinfo  {journal} {Computers and Fluids}\ }\textbf {\bibinfo
  {volume} {192}} (\bibinfo {year} {2019}),\
  10.1016/j.compfluid.2019.104258}\BibitemShut {NoStop}%
\bibitem [{\citenamefont {Brener}\ \emph {et~al.}(2021)\citenamefont {Brener},
  \citenamefont {Cruz}, \citenamefont {Thompson},\ and\ \citenamefont
  {Anjos}}]{Brener2021}%
  \BibitemOpen
  \bibfield  {author} {\bibinfo {author} {\bibfnamefont {B.~P.}\ \bibnamefont
  {Brener}}, \bibinfo {author} {\bibfnamefont {M.~A.}\ \bibnamefont {Cruz}},
  \bibinfo {author} {\bibfnamefont {R.~L.}\ \bibnamefont {Thompson}}, \ and\
  \bibinfo {author} {\bibfnamefont {R.~P.}\ \bibnamefont {Anjos}},\ }\bibfield
  {title} {\enquote {\bibinfo {title} {{Conditioning and accurate solutions of
  Reynolds average Navier-Stokes equations with data-driven turbulence
  closures}},}\ }\href {\doibase 10.1017/jfm.2021.148} {\bibfield  {journal}
  {\bibinfo  {journal} {Journal of Fluid Mechanics}\ }\textbf {\bibinfo
  {volume} {915}},\ \bibinfo {pages} {1--27} (\bibinfo {year}
  {2021})}\BibitemShut {NoStop}%
\bibitem [{\citenamefont {Wu}\ \emph {et~al.}(2019)\citenamefont {Wu},
  \citenamefont {Sun}, \citenamefont {Laizet},\ and\ \citenamefont
  {Xiao}}]{Wu2019}%
  \BibitemOpen
  \bibfield  {author} {\bibinfo {author} {\bibfnamefont {J.~L.}\ \bibnamefont
  {Wu}}, \bibinfo {author} {\bibfnamefont {R.}~\bibnamefont {Sun}}, \bibinfo
  {author} {\bibfnamefont {S.}~\bibnamefont {Laizet}}, \ and\ \bibinfo {author}
  {\bibfnamefont {H.}~\bibnamefont {Xiao}},\ }\bibfield  {title} {\enquote
  {\bibinfo {title} {{Representation of stress tensor perturbations with
  application in machine-learning-assisted turbulence modeling}},}\ }\href
  {\doibase 10.1016/j.cma.2018.09.010} {\bibfield  {journal} {\bibinfo
  {journal} {Computer Methods in Applied Mechanics and Engineering}\ }\textbf
  {\bibinfo {volume} {346}},\ \bibinfo {pages} {707--726} (\bibinfo {year}
  {2019})}\BibitemShut {NoStop}%
\bibitem [{\citenamefont {Hanjali{\'{c}}}, \citenamefont {Popovac},\ and\
  \citenamefont {Had{\v{z}}iabdi{\'{c}}}(2004)}]{Hanjalic2004}%
  \BibitemOpen
  \bibfield  {author} {\bibinfo {author} {\bibfnamefont {K.}~\bibnamefont
  {Hanjali{\'{c}}}}, \bibinfo {author} {\bibfnamefont {M.}~\bibnamefont
  {Popovac}}, \ and\ \bibinfo {author} {\bibfnamefont {M.}~\bibnamefont
  {Had{\v{z}}iabdi{\'{c}}}},\ }\bibfield  {title} {\enquote {\bibinfo {title}
  {{A robust near-wall elliptic-relaxation eddy-viscosity turbulence model for
  CFD}},}\ }\href {\doibase 10.1016/j.ijheatfluidflow.2004.07.005} {\bibfield
  {journal} {\bibinfo  {journal} {International Journal of Heat and Fluid
  Flow}\ }\textbf {\bibinfo {volume} {25}},\ \bibinfo {pages} {1047--1051}
  (\bibinfo {year} {2004})}\BibitemShut {NoStop}%
\bibitem [{\citenamefont {Durbin}(1991)}]{Durbin1991}%
  \BibitemOpen
  \bibfield  {author} {\bibinfo {author} {\bibfnamefont {P.~A.}\ \bibnamefont
  {Durbin}},\ }\bibfield  {title} {\enquote {\bibinfo {title} {{Near-wall
  turbulence closure modeling without "damping functions"}},}\ }\href {\doibase
  10.1007/BF00271513} {\bibfield  {journal} {\bibinfo  {journal} {Theoretical
  and Computational Fluid Dynamics}\ }\textbf {\bibinfo {volume} {3}},\
  \bibinfo {pages} {1--13} (\bibinfo {year} {1991})}\BibitemShut {NoStop}%
\bibitem [{\citenamefont {McConkey}, \citenamefont {Yee},\ and\ \citenamefont
  {Lien}(2021)}]{McConkey2021b}%
  \BibitemOpen
  \bibfield  {author} {\bibinfo {author} {\bibfnamefont {R.}~\bibnamefont
  {McConkey}}, \bibinfo {author} {\bibfnamefont {E.}~\bibnamefont {Yee}}, \
  and\ \bibinfo {author} {\bibfnamefont {F.~S.}\ \bibnamefont {Lien}},\
  }\bibfield  {title} {\enquote {\bibinfo {title} {{A curated dataset for
  data-driven turbulence modelling}},}\ }\href {\doibase
  10.1038/s41597-021-01034-2} {\bibfield  {journal} {\bibinfo  {journal}
  {Scientific Data}\ }\textbf {\bibinfo {volume} {8}},\ \bibinfo {pages}
  {1--14} (\bibinfo {year} {2021})}\BibitemShut {NoStop}%
\bibitem [{\citenamefont {Xiao}\ \emph {et~al.}(2020)\citenamefont {Xiao},
  \citenamefont {Wu}, \citenamefont {Laizet},\ and\ \citenamefont
  {Duan}}]{Xiao2020}%
  \BibitemOpen
  \bibfield  {author} {\bibinfo {author} {\bibfnamefont {H.}~\bibnamefont
  {Xiao}}, \bibinfo {author} {\bibfnamefont {J.~L.}\ \bibnamefont {Wu}},
  \bibinfo {author} {\bibfnamefont {S.}~\bibnamefont {Laizet}}, \ and\ \bibinfo
  {author} {\bibfnamefont {L.}~\bibnamefont {Duan}},\ }\bibfield  {title}
  {\enquote {\bibinfo {title} {{Flows over periodic hills of parameterized
  geometries: A dataset for data-driven turbulence modeling from direct
  simulations}},}\ }\href {\doibase 10.1016/j.compfluid.2020.104431} {\bibfield
   {journal} {\bibinfo  {journal} {Computers and Fluids}\ }\textbf {\bibinfo
  {volume} {200}},\ \bibinfo {pages} {104431} (\bibinfo {year}
  {2020})}\BibitemShut {NoStop}%
\bibitem [{\citenamefont {Thompson}, \citenamefont {Mompean},\ and\
  \citenamefont {Laurent}(2010)}]{Thompson2010}%
  \BibitemOpen
  \bibfield  {author} {\bibinfo {author} {\bibfnamefont {R.~L.}\ \bibnamefont
  {Thompson}}, \bibinfo {author} {\bibfnamefont {G.}~\bibnamefont {Mompean}}, \
  and\ \bibinfo {author} {\bibfnamefont {T.}~\bibnamefont {Laurent}},\
  }\bibfield  {title} {\enquote {\bibinfo {title} {{A methodology to quantify
  the nonlinearity of the Reynolds stress tensor}},}\ }\href {\doibase
  10.1080/14685248.2010.501799} {\bibfield  {journal} {\bibinfo  {journal}
  {Journal of Turbulence}\ }\textbf {\bibinfo {volume} {11}},\ \bibinfo {pages}
  {1--27} (\bibinfo {year} {2010})}\BibitemShut {NoStop}%
\bibitem [{\citenamefont {Pope}(2000)}]{Pope2000}%
  \BibitemOpen
  \bibfield  {author} {\bibinfo {author} {\bibfnamefont {S.~B.}\ \bibnamefont
  {Pope}},\ }\href@noop {} {\emph {\bibinfo {title} {{Turbulent Flows}}}}\
  (\bibinfo  {publisher} {Cambridge University Press},\ \bibinfo {address}
  {Cambridge, UK},\ \bibinfo {year} {2000})\BibitemShut {NoStop}%
\bibitem [{\citenamefont {Ho}\ and\ \citenamefont {West}(2021)}]{Ho2021}%
  \BibitemOpen
  \bibfield  {author} {\bibinfo {author} {\bibfnamefont {J.}~\bibnamefont
  {Ho}}\ and\ \bibinfo {author} {\bibfnamefont {A.}~\bibnamefont {West}},\
  }\bibfield  {title} {\enquote {\bibinfo {title} {{Field Inversion and Machine
  Learning for turbulence modelling applied to three-dimensional separated
  flows}},}\ }\href {\doibase 10.2514/6.2021-2903} {\  (\bibinfo {year}
  {2021}),\ 10.2514/6.2021-2903}\BibitemShut {NoStop}%
\bibitem [{\citenamefont {Laurence}, \citenamefont {Uribe},\ and\ \citenamefont
  {Utyuzhnikov}(2005)}]{Laurence2005}%
  \BibitemOpen
  \bibfield  {author} {\bibinfo {author} {\bibfnamefont {D.~R.}\ \bibnamefont
  {Laurence}}, \bibinfo {author} {\bibfnamefont {J.~C.}\ \bibnamefont {Uribe}},
  \ and\ \bibinfo {author} {\bibfnamefont {S.~V.}\ \bibnamefont
  {Utyuzhnikov}},\ }\bibfield  {title} {\enquote {\bibinfo {title} {{A robust
  formulation of the v2-f model}},}\ }\href {\doibase
  10.1007/s10494-005-1974-8} {\bibfield  {journal} {\bibinfo  {journal} {Flow,
  Turbulence and Combustion}\ }\textbf {\bibinfo {volume} {73}},\ \bibinfo
  {pages} {169--185} (\bibinfo {year} {2005})}\BibitemShut {NoStop}%
\bibitem [{\citenamefont {{OpenCFD Ltd.}}(2019)}]{Openfoam}%
  \BibitemOpen
  \bibfield  {author} {\bibinfo {author} {\bibnamefont {{OpenCFD Ltd.}}},\
  }\href {https://www.openfoam.com/documentation/guides/latest/doc/} {\enquote
  {\bibinfo {title} {{OpenFOAM: User Guide v2006}},}\ } (\bibinfo {year}
  {2019})\BibitemShut {NoStop}%
\bibitem [{\citenamefont {Lien}\ and\ \citenamefont
  {Kalitzin}(2001)}]{Lien2001}%
  \BibitemOpen
  \bibfield  {author} {\bibinfo {author} {\bibfnamefont {F.~S.}\ \bibnamefont
  {Lien}}\ and\ \bibinfo {author} {\bibfnamefont {G.}~\bibnamefont
  {Kalitzin}},\ }\bibfield  {title} {\enquote {\bibinfo {title} {{Computations
  of transonic flow with the v2-f turbulence model}},}\ }\href {\doibase
  10.1016/S0142-727X(00)00073-4} {\bibfield  {journal} {\bibinfo  {journal}
  {International Journal of Heat and Fluid Flow}\ }\textbf {\bibinfo {volume}
  {22}},\ \bibinfo {pages} {53--61} (\bibinfo {year} {2001})}\BibitemShut
  {NoStop}%
\bibitem [{\citenamefont {Klambauer}\ \emph {et~al.}(2017)\citenamefont
  {Klambauer}, \citenamefont {Unterthiner}, \citenamefont {Mayr},\ and\
  \citenamefont {Hochreiter}}]{Klambauer2017}%
  \BibitemOpen
  \bibfield  {author} {\bibinfo {author} {\bibfnamefont {G.}~\bibnamefont
  {Klambauer}}, \bibinfo {author} {\bibfnamefont {T.}~\bibnamefont
  {Unterthiner}}, \bibinfo {author} {\bibfnamefont {A.}~\bibnamefont {Mayr}}, \
  and\ \bibinfo {author} {\bibfnamefont {S.}~\bibnamefont {Hochreiter}},\
  }\bibfield  {title} {\enquote {\bibinfo {title} {{Self-normalizing neural
  networks}},}\ }\href@noop {} {\bibfield  {journal} {\bibinfo  {journal}
  {Advances in Neural Information Processing Systems (NIPS 2017)}\ }\textbf
  {\bibinfo {volume} {2017-Decem}},\ \bibinfo {pages} {972--981} (\bibinfo
  {year} {2017})}\BibitemShut {NoStop}%
\bibitem [{\citenamefont {G\'eron}(2019)}]{Geron}%
  \BibitemOpen
  \bibfield  {author} {\bibinfo {author} {\bibfnamefont {A.}~\bibnamefont
  {G\'eron}},\ }\href@noop {} {\emph {\bibinfo {title} {{Hands-On Machine
  Learning with Scikit-Learn, Keras \& Tensorflow}}}},\ \bibinfo {edition}
  {2nd}\ ed.\ (\bibinfo  {publisher} {O'Reilly Media Inc.},\ \bibinfo {address}
  {Sebastopol, CA},\ \bibinfo {year} {2019})\ p.\ \bibinfo {pages}
  {1150}\BibitemShut {NoStop}%
\bibitem [{\citenamefont {Chollet}\ and\ \citenamefont
  {{others}}(2015)}]{keras}%
  \BibitemOpen
  \bibfield  {author} {\bibinfo {author} {\bibfnamefont {F.}~\bibnamefont
  {Chollet}}\ and\ \bibinfo {author} {\bibnamefont {{others}}},\ }\href@noop {}
  {\enquote {\bibinfo {title} {{Keras}},}\ }\bibinfo {howpublished}
  {{\textbackslash}url{\{}https://keras.io{\}}} (\bibinfo {year}
  {2015})\BibitemShut {NoStop}%
\bibitem [{\citenamefont {Spencer}\ and\ \citenamefont
  {Rivlin}(1962)}]{Spencer1962}%
  \BibitemOpen
  \bibfield  {author} {\bibinfo {author} {\bibfnamefont {A.~J.}\ \bibnamefont
  {Spencer}}\ and\ \bibinfo {author} {\bibfnamefont {R.~S.}\ \bibnamefont
  {Rivlin}},\ }\bibfield  {title} {\enquote {\bibinfo {title} {{Isotropic
  integrity bases for vectors and second-order tensors - Part I}},}\ }\href
  {\doibase 10.1007/BF00253332} {\bibfield  {journal} {\bibinfo  {journal}
  {Archive for Rational Mechanics and Analysis}\ }\textbf {\bibinfo {volume}
  {9}},\ \bibinfo {pages} {45--63} (\bibinfo {year} {1962})}\BibitemShut
  {NoStop}%
\bibitem [{\citenamefont {Pedregosa}\ \emph {et~al.}(2011)\citenamefont
  {Pedregosa}, \citenamefont {Varoquaux}, \citenamefont {Gramfort},
  \citenamefont {Michel}, \citenamefont {Thirion}, \citenamefont {Grisel},
  \citenamefont {Blondel}, \citenamefont {Prettenhofer}, \citenamefont {Weiss},
  \citenamefont {Dubourg}, \citenamefont {Vanderplas}, \citenamefont {Passos},
  \citenamefont {Cournapeau}, \citenamefont {Brucher}, \citenamefont {Perrot},\
  and\ \citenamefont {Duchesnay}}]{scikit-learn}%
  \BibitemOpen
  \bibfield  {author} {\bibinfo {author} {\bibfnamefont {F.}~\bibnamefont
  {Pedregosa}}, \bibinfo {author} {\bibfnamefont {G.}~\bibnamefont
  {Varoquaux}}, \bibinfo {author} {\bibfnamefont {A.}~\bibnamefont {Gramfort}},
  \bibinfo {author} {\bibfnamefont {V.}~\bibnamefont {Michel}}, \bibinfo
  {author} {\bibfnamefont {B.}~\bibnamefont {Thirion}}, \bibinfo {author}
  {\bibfnamefont {O.}~\bibnamefont {Grisel}}, \bibinfo {author} {\bibfnamefont
  {M.}~\bibnamefont {Blondel}}, \bibinfo {author} {\bibfnamefont
  {P.}~\bibnamefont {Prettenhofer}}, \bibinfo {author} {\bibfnamefont
  {R.}~\bibnamefont {Weiss}}, \bibinfo {author} {\bibfnamefont
  {V.}~\bibnamefont {Dubourg}}, \bibinfo {author} {\bibfnamefont
  {J.}~\bibnamefont {Vanderplas}}, \bibinfo {author} {\bibfnamefont
  {A.}~\bibnamefont {Passos}}, \bibinfo {author} {\bibfnamefont
  {D.}~\bibnamefont {Cournapeau}}, \bibinfo {author} {\bibfnamefont
  {M.}~\bibnamefont {Brucher}}, \bibinfo {author} {\bibfnamefont
  {M.}~\bibnamefont {Perrot}}, \ and\ \bibinfo {author} {\bibfnamefont
  {E.}~\bibnamefont {Duchesnay}},\ }\bibfield  {title} {\enquote {\bibinfo
  {title} {{Scikit-learn: Machine Learning in Python}},}\ }\href@noop {}
  {\bibfield  {journal} {\bibinfo  {journal} {Journal of Machine Learning
  Research}\ }\textbf {\bibinfo {volume} {12}},\ \bibinfo {pages} {2825--2830}
  (\bibinfo {year} {2011})}\BibitemShut {NoStop}%
\bibitem [{\citenamefont {Lundberg}\ and\ \citenamefont
  {Lee}(2017)}]{Lundberg2017}%
  \BibitemOpen
  \bibfield  {author} {\bibinfo {author} {\bibfnamefont {S.~M.}\ \bibnamefont
  {Lundberg}}\ and\ \bibinfo {author} {\bibfnamefont {S.-I.}\ \bibnamefont
  {Lee}},\ }\bibfield  {title} {\enquote {\bibinfo {title} {{A Unified Approach
  to Interpreting Model Predictions}},}\ }in\ \href
  {https://proceedings.neurips.cc/paper/2017/file/8a20a8621978632d76c43dfd28b67767-Paper.pdf}
  {\emph {\bibinfo {booktitle} {Advances in Neural Information Processing
  Systems}}},\ Vol.~\bibinfo {volume} {30},\ \bibinfo {editor} {edited by\
  \bibinfo {editor} {\bibfnamefont {I.}~\bibnamefont {Guyon}}, \bibinfo
  {editor} {\bibfnamefont {U.~V.}\ \bibnamefont {Luxburg}}, \bibinfo {editor}
  {\bibfnamefont {S.}~\bibnamefont {Bengio}}, \bibinfo {editor} {\bibfnamefont
  {H.}~\bibnamefont {Wallach}}, \bibinfo {editor} {\bibfnamefont
  {R.}~\bibnamefont {Fergus}}, \bibinfo {editor} {\bibfnamefont
  {S.}~\bibnamefont {Vishwanathan}}, \ and\ \bibinfo {editor} {\bibfnamefont
  {R.}~\bibnamefont {Garnett}}}\ (\bibinfo  {publisher} {Curran Associates,
  Inc.},\ \bibinfo {year} {2017})\BibitemShut {NoStop}%
\bibitem [{\citenamefont {Lardeau}\ and\ \citenamefont
  {Billard}(2016)}]{Lardeau2016}%
  \BibitemOpen
  \bibfield  {author} {\bibinfo {author} {\bibfnamefont {S.}~\bibnamefont
  {Lardeau}}\ and\ \bibinfo {author} {\bibfnamefont {F.}~\bibnamefont
  {Billard}},\ }\href@noop {} {\enquote {\bibinfo {title} {Development of an
  elliptic-blending lag model for industrial applications},}\ }\bibinfo
  {howpublished} {Presented at AIAA Aerospace Science Meeting, 54th, Jan. 4--8,
  San Diego, CA, AIAA Paper 2005--522} (\bibinfo {year} {2016})\BibitemShut
  {NoStop}%
\end{thebibliography}%

\end{document}